\documentclass[a4paper,fleqn,usenatbib,useAMS]{mnras}
\pdfoutput=1

\usepackage{graphicx}	
\usepackage{amsmath}	
\usepackage{amssymb}	
\usepackage{multicol}        
\usepackage{bm}		
\usepackage{pdflscape}	
\usepackage[T1]{fontenc}
\usepackage{aecompl}

\usepackage[T1]{fontenc}
\usepackage{ae,aecompl}
\usepackage{natbib}
\usepackage[export]{adjustbox}

\usepackage{newtxtext,newtxmath}

\title[Bulge effect on bar instability]{A Study of the Effect of Bulges on Bar Formation in Disk galaxies }

\author[Kataria \& Das]{Sandeep Kumar Kataria $^{1,2}$\thanks{Contact e-mail: \href{mailto:sandeep.kataria@iiap.res.in}{sandeep.kataria@iiap.res.in}} Mousumi Das $^1$\thanks{Contact e-mail: \href{mailto:mousumi@iiap.res.in}{mousumi@iiap.res.in}}
\\
$^{1}$Indian Institute of Astrophysics, Koramangla, Bangalore-560034\\
$^{2}$Indian Institute of Science, Bangalore-560012}

\pubyear{2016}

\begin{document}
\label{firstpage}
\pagerange{\pageref{firstpage}--\pageref{lastpage}}
\maketitle

\begin{abstract}
We use N-body simulations of bar formation in isolated galaxies to study the effect of bulge mass and bulge concentration on bar formation. Bars are global disk instabilities that evolve by transferring angular momentum from the inner to outer disks and to the dark matter halo. It is well known that a massive spherical component such as halo in a disk galaxy can make it bar stable. In this study we explore the effect of another spherical component, the bulge, on bar formation in disk galaxies. In our models we vary both the bulge mass and concentration. We have used two sets of models, one that has a dense bulge and high surface density disk. The second model has a less concentrated bulge and a lighter disk. In both models we vary the bulge to disk mass fraction from 0 to 0.7. Simulations of both the models show that there is an upper cutoff in bulge to disk mass ratio $M_b/M_d$ above which bars cannot form; the cutoff is smaller for denser bulges( $M_b/M_d = 0.2$) compared to less denser ones ($M_b/M_d = 0.5$). We define a new criteria for bar formation in terms of  bulge to total radial force ratio ($F_b/F_{tot}$) at the disk scale lengths above which bars cannot form. We find that if  $F_b/F_{tot}~$>~0.35, a disk is stable and a bar cannot form. Our results indicate that early type disk galaxies can still form strong bars in spite of having massive bulges.
\end{abstract}

\begin{keywords}
galaxies:bulges-galaxies:kinematics and dynamics-galaxies:evolution-galaxies:structure-methods: numerical-cosmology:dark matter
\end{keywords}



\section{Introduction}
Nearly two-thirds of all disk galaxies in the observable universe are barred \citep{3,4a,1,2,4}. Many studies show that the fraction of barred galaxies decreases with redshift \citep{22,23,melvin.etal.2014} but some observations have found constant bar fractions upto z$\sim$1 \citep{5,5a,6} or even z$\sim$2 \citep{simmons.etal.2014} which may be attributed to inclusion of small size bars and different galaxy selection criteria in different surveys. However, it is clear from these observations that most bars survive for at least t$\sim$8~Gyr in galaxies in our low redshift Universe. \\

In the Hubble sequence bars vary from early to late type spiral galaxies \citep{1998gaas.book.....B,6a,6b}. The bars associated with early type, bulge dominated galaxies (SBa, SBb) appear to be longer than those found in late type spirals (Sc) \citep{erwin.2005}. There is also a significant coorelation between bar strength and bulge mass \citep{31,30a,32}.
The bulges themselves can be broadly classified into two types, the classical bulges that have a sersic index $n~>~2$ and the more oval or disky pseudobulges that have $n~<~2$ \citep{fisher.drory.2008}.
Their structures suggests different formation mechanisms for each of them \citep{gadotti.etal.2009}. The origin of classical bulges is thought to be a result of major mergers \citep{35,36,37,38}, multiple minor mergers \citep{39,40}, monolithic collapse of primordial gas clouds \citep{41} or the accretion of small satellites \citep{42}. Pseudo bulges are formed due to disk instability during secular evolution \citep{42b,42c} or buckling instability of bars \citep{combes.1990, raha.etal.1991,Martinez.2006}, inward pull of gas by bars \citep{13a} and heating of bars by vertical resonances \citep{33}, resulting in boxy/peanut shaped bulges.

Theoretical studies have shown that bars will form in self gravitating, rotating disks when most of the kinetic energy is in rotational motion i.e. in cold disks \citep{12a,12b} unless there is a massive dark matter halo that stabilizes it against bar formation by providing a spherical gravitational field \citep{24}. However recent studies \citep{14a,15a1} show that a cold disk can become bar unstable despite the presence of  a massive halo, which shakes the credibility of these previous studies. Global disk instabilities such as bars and spirals arms can also be triggered by disk perturbations due to interactions/mergers with minor satellites or larger galaxies \citep{7,8,9,10,11,12}. The evolution of bars in disk galaxies has been extensively studied with N-body simulations over the past few decades \citep{13,13a,14,14a,15,15a1,15a,29,17,18}. The disk-halo interaction has been shown to play an important role not only in bar formation but also bar rotation speeds \citep{2007MNRAS.375..460W,0004-637X-543-2-704}. Studies with live halos show that bars evolve secularly in isolated galaxies through the exchange of angular momentum between disk stars and halo particles at radii corresponding to the disk resonances \citep{18}. Faster rotating halos may also trigger early bar formation \citep{18} but can also make bars weaker in the secular evolution phase. The bulges also gain angular momentum from the disk at resonances and this results in the spinning up of bulges\citep{51}. These studies are similar to previous classical studies in which disk stars lose angular momentum at resonances \citep{19,20} or by being trapped in slowly rotating bar structures through dynamical friction \citep{21}. 

Although there have been many simulation studies of the effect of disk-halo interaction on bars, there are very few studies of the effect of bulges on bars. Since bulges contribute significantly to the radial force in disks, bulge masses and sizes can affect bar formation \citep{45}. One of the earlier theoretical studies of the effect of bulges on disk instabilities \citep{44} showed that the presence of a strong bulge component in disk galaxies cuts off the feedback mechanism during swing amplification \citep{43} by introducing an Inner Lindblad Resonace (ILR) in the disk. The ILR does not allow trailing waves to go through the center and emerge out as leading waves; this prohibits the growth of bar type instabilities. However, detailed N-body simulations showed that this is not always true \citep{46}; about 50 to 70 $\%$ of total galactic baryonic mass has to be in a spherical component to prevent the onset of bar instability in disk galaxies. These early studies used a spherical mass smaller than the disk but referred to it as a "halo" rather than a bulge \citep{46}. Further studies \citep{47} showed that bar type instabilities can be sustained irrespective of the density of the bulge component. However, these studies used rigid bulges \citep{46,47}. When live bulges were used, it was found that the bulges gained angular momentum from the disks and this increased as the bar evolved \citep{46}. The disadvantage of all these early studies is that they did not include dark matter halo components in their galaxy models.

The above discussion holds for mainly isolated galaxies that are secularly evolving. However, when there are external perturbations to galaxy disks due to interactions with nearby galaxies, stellar bars can form even in the presence of a strong bulge. In such cases the ILR is not able to prevent the growth of bars. Later studies that included halo components in their models suggested that the presence of massive bulges make disks stable against bar formation \citep{44a}. It is well known that the probability of having an ILR is higher when there is a massive bulge as the angular velocity $\Omega$ of stars in the disk increases and hence the critical value $\Omega_{crtical}~=~\Omega_{p} + \kappa/2 $ will be reached, therefore cutting off feedback mechanism as discussed in the last paragraph, where $\Omega_{p}$ and $\kappa$ are the pattern speed of bar and epicyclic frequency of stars. However, there are many non-linear processes in galaxy evolution which allow bar formation in bulge dominated disk galaxies \citep{49,50}. Cosmological hydrodynamic simulations \citep{48} also show that strong bars can form in galaxies that have prominent bulges rather than those without bulges.

Recent studies have shown that bars can also affect bulges by increasing their spin \citep{51,52,53}. However, there is no detailed study which explores the effect of bulge masses on bar formation and its evolution. In this paper we re-visit the dependence of bar formation on bulges but with a difference; we vary not only the bulge mass but also its concentration.  We also focus on the angular momentum transfer between the disk and bulge, since angular momentum transfer between disks and dark matter halos has been found to play a key role in bar evolution \citep{18}. Thus the goal of this paper is to determine how bulge mass and concentration affects bar formation, morphology, pattern speed and angular momentum transfer between the different galaxy components (bulge, disk, halo). Our models start from bulgeless galaxies to bulge dominated disks where the bulge to disk mass ratio can be as high as 70\%. We have two sets of models, one with a concentrated bulge and the other with a less concentrated bulge. The plan of the paper is as follows. In Section~\ref{Technique} we discuss the numerical techniques used for this work which includes initial condition generation and simulation methods. In Section~\ref{Results} we discuss the evolution of various bar parameters in the different cases, such as pattern speed, bar strength, angular momentum transfer between disk and bulge components and the origin of pseudo bulges in our models. Apart from this in the same section we discuss a model independent parameter that can be used to determine the limiting bulge force for bar formation. In Section~\ref{Discussion} we discuss the implication of our results for bulge-bar correlations in disk galaxies. Finally in the Section~\ref{summary} we summarize our work.

\section{Numerical Technique}\label{Technique}
\subsection{Initial Conditions of modelled galaxies} 
For generating initial disk galaxy models we have used the code GalIC \citep{55} which iteratively populates orbits with given density distribution using the parts of Swartzschild technique. This code finds steady state solution for the collisionless boltzman equation by iteratively adjusting the  velocity of disk stars in a given density distribution to generate equilibrium galaxy models. The number of particles used for this simulation is $10^6$ dark matter halo particles, $10^5$ disk particles and 5x$10^4$ bulge particles. 

\begin{figure}
\includegraphics[scale=0.437]{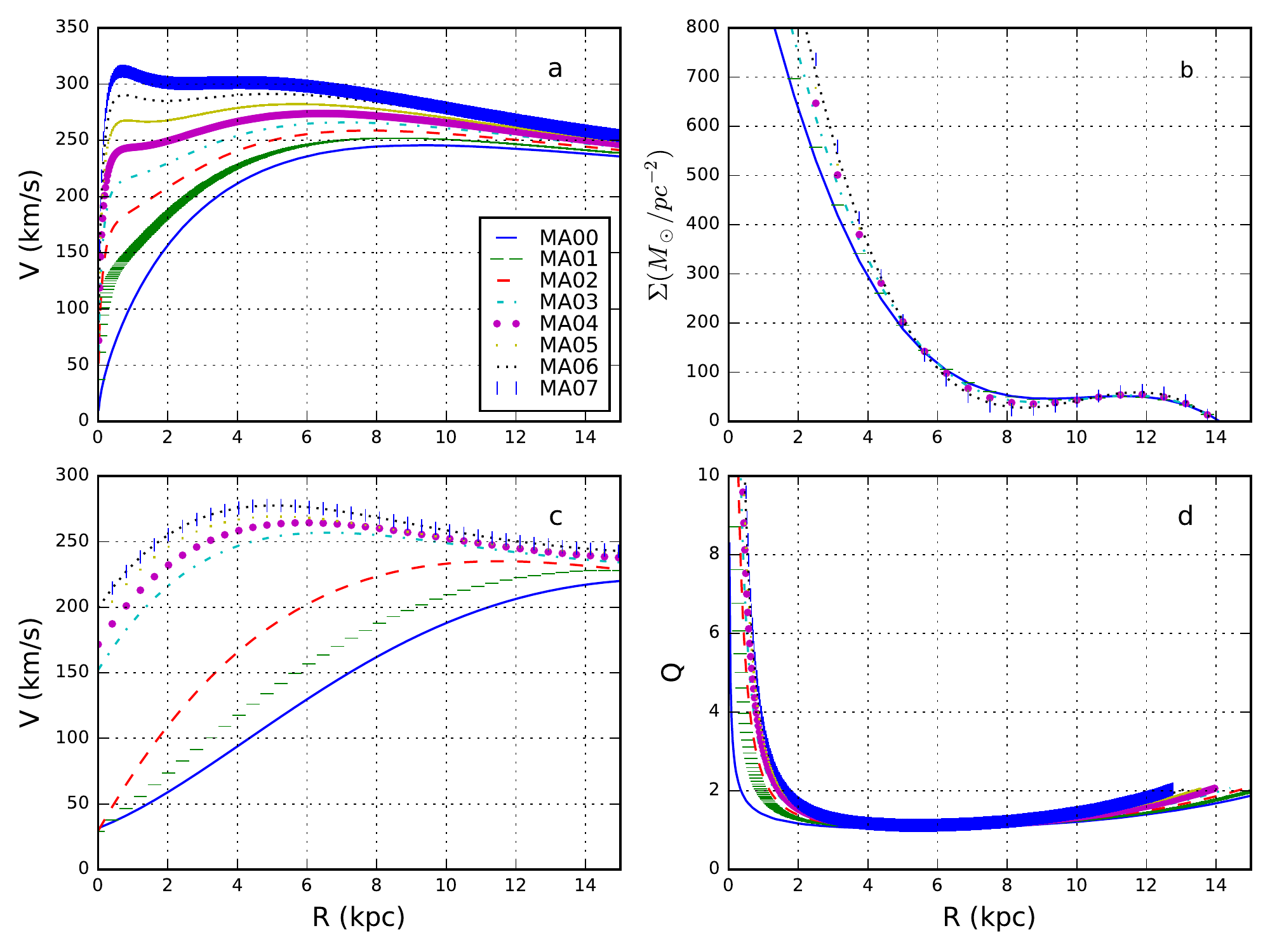}
\caption{a)Initial rotation curves of disk stars; b)Surface density; c)final rotation curve at 9.78Gyr. d)toomre's parameter variation with radius for all the MA models}
\label{fig rot_MA}
\end{figure}

\begin{figure}
\includegraphics[scale=0.4]{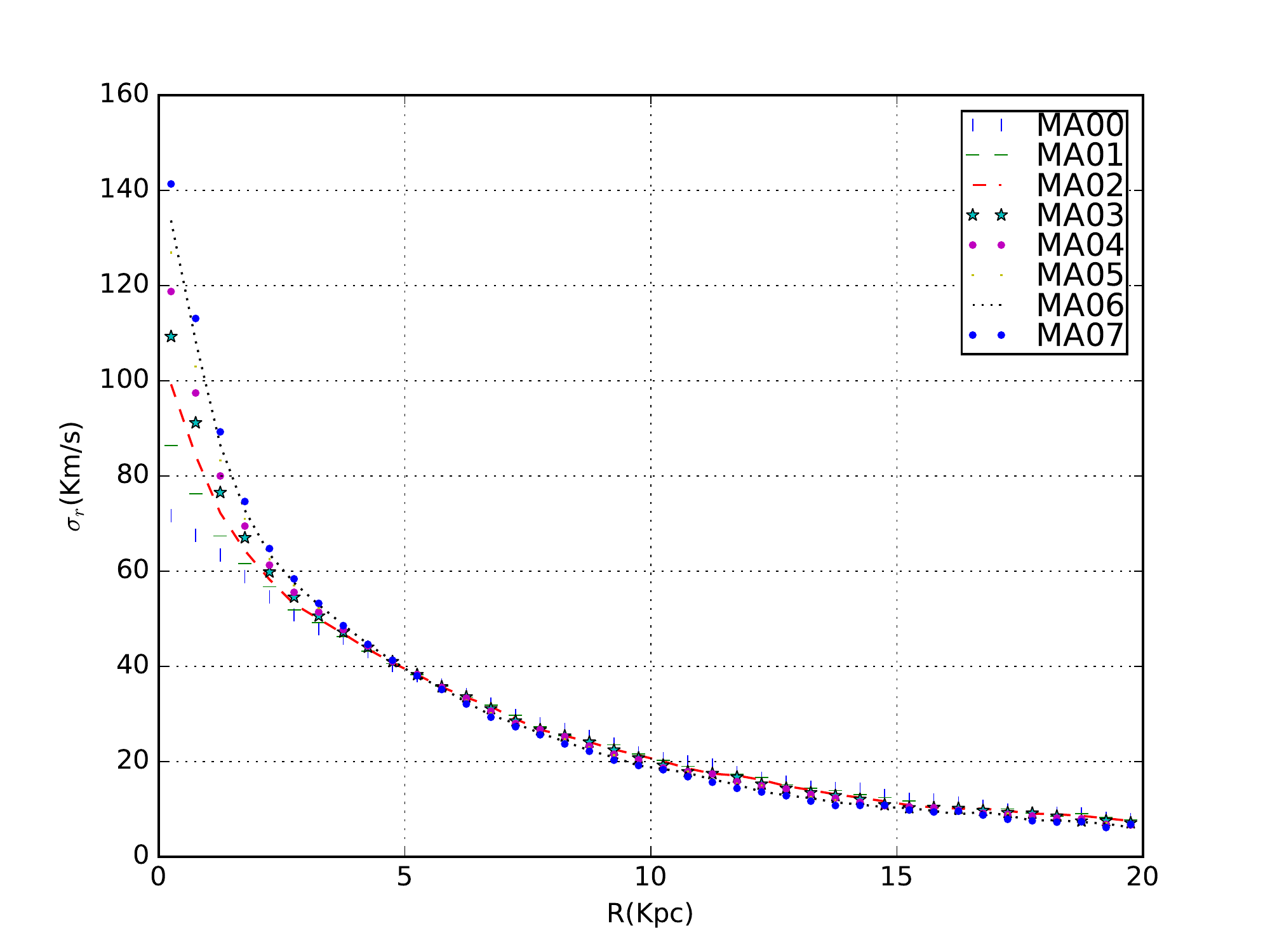}
\caption{Initial radial velocity dispersions for all the MA models}
\label{fig sigma_MA}
\end{figure}

\begin{figure}

\includegraphics[scale=0.47]{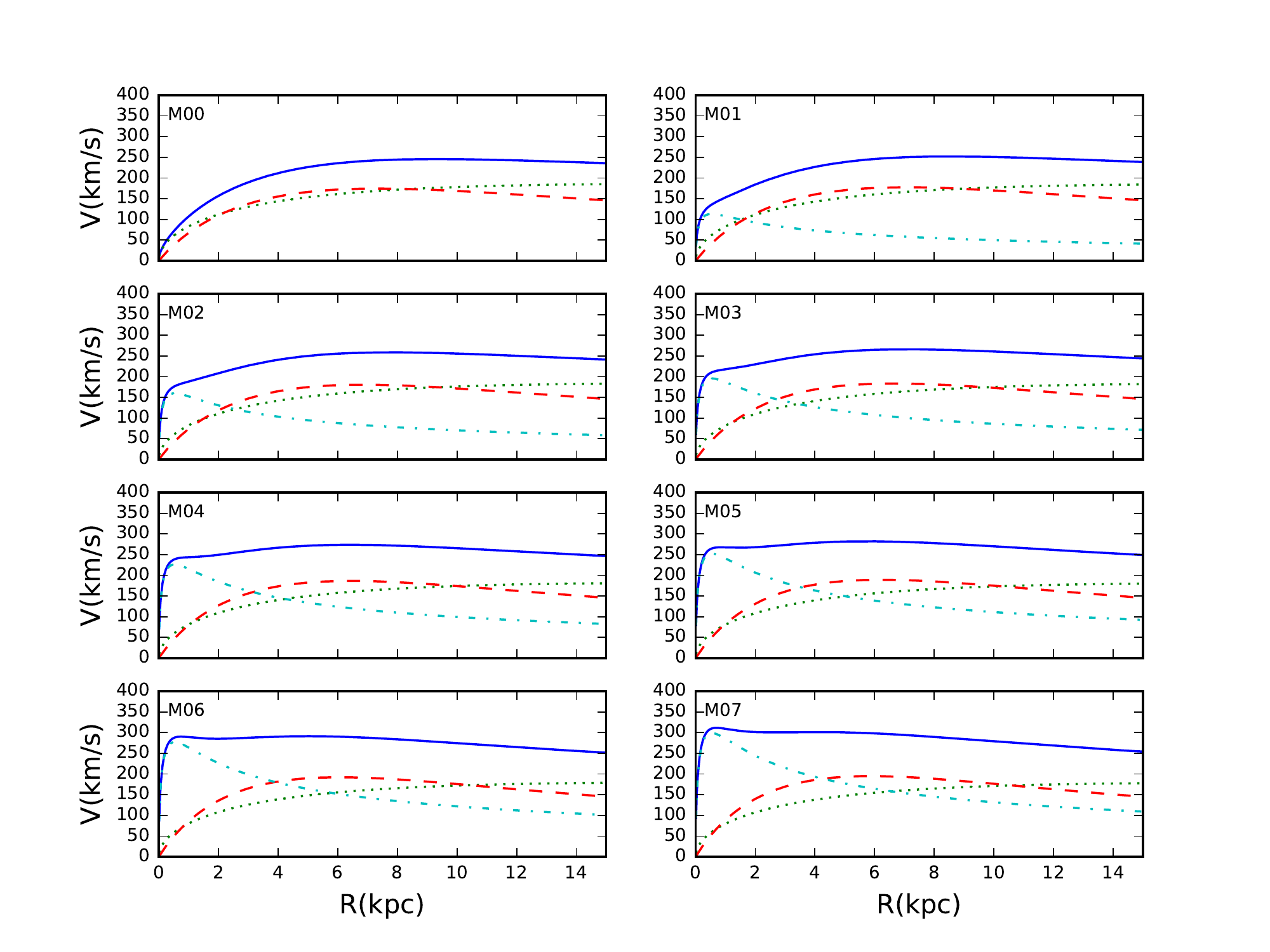}
\caption{Represents rotational curves for all of our MA models. Here solid line shows rotational velocity due to all the components; Halo (dotted line);disk (dashed line);Bulge (dash-dot line)}
\label{fig:rotcomp_MA}
\end{figure}

\begin{figure}
\includegraphics[scale=0.45]{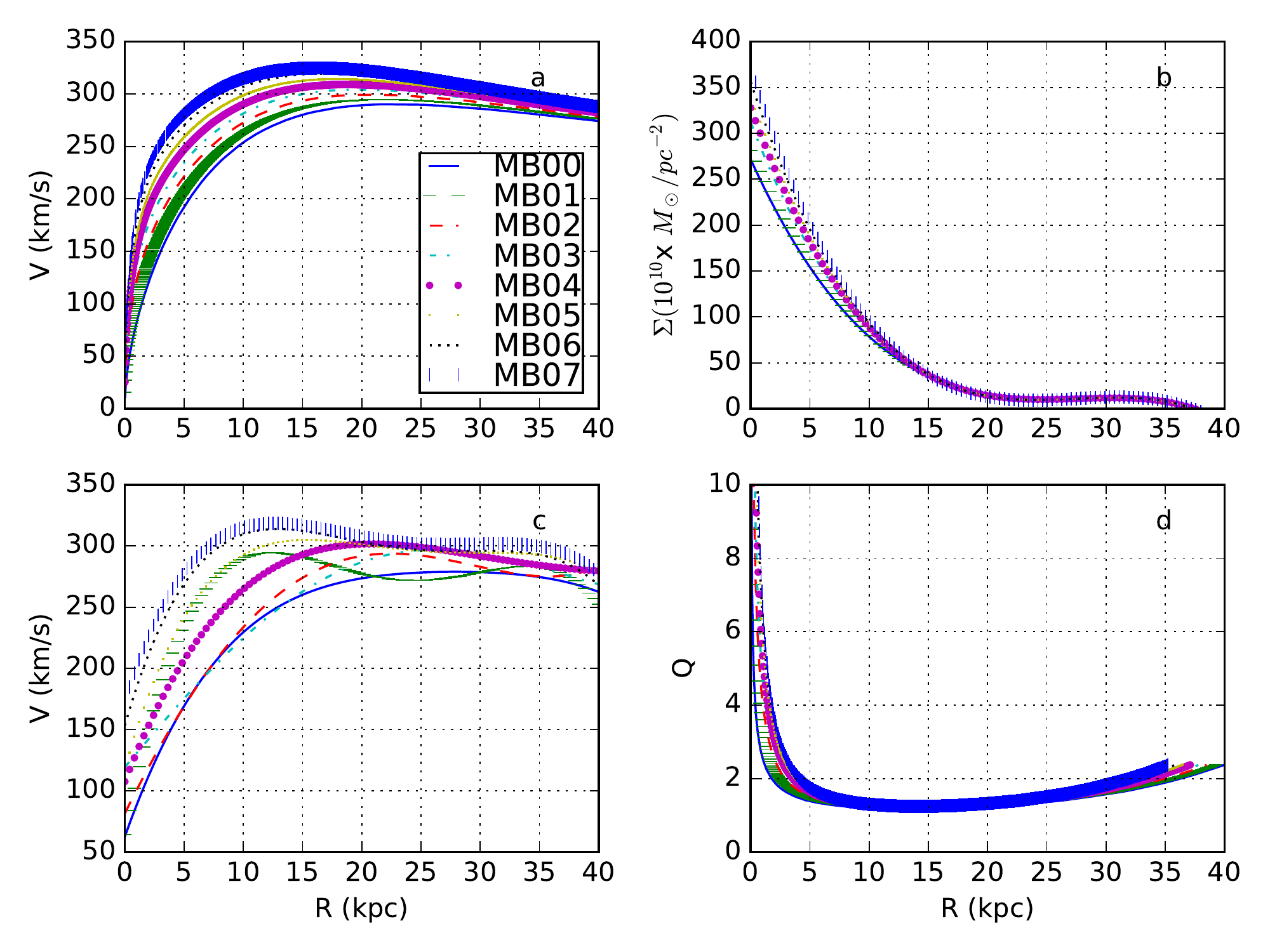}
\caption{a)Initial rotation curves of disk stars; b)surface density; c)final rotation curve at 9.78Gyr; d)toomre's parameter variation with radius for all the MB models}
\label{fig rot_MB}
\end{figure}

\begin{figure}
\includegraphics[scale=0.47]{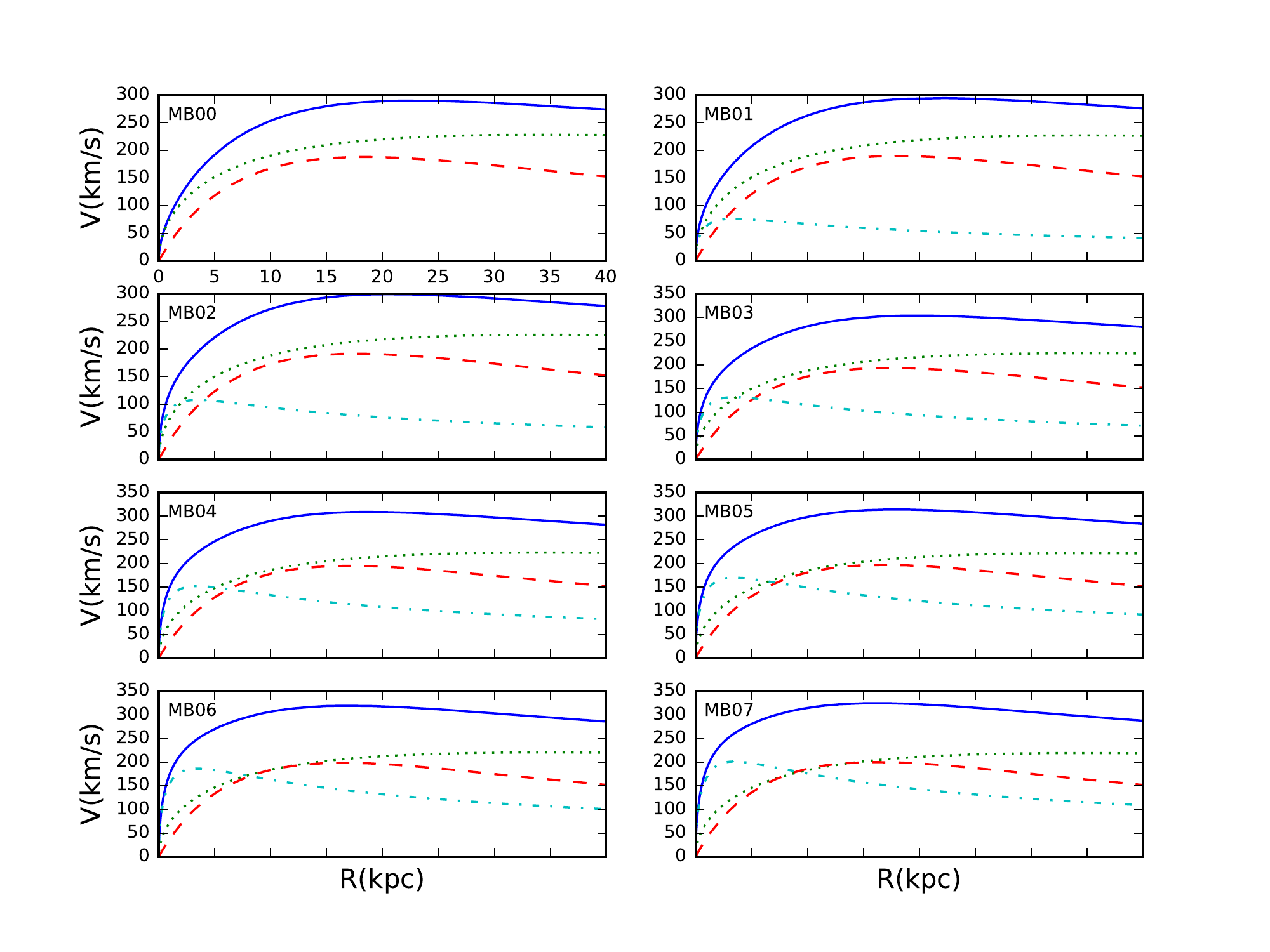}
\caption{Represents rotational curves for all of our MB models. Here solid line shows rotational velocity due to all the components; Halo (dotted line);disk (dashed line);Bulge (dash-dot line)}
\label{fig:rotcomp_MB}
\end{figure}

\begin{figure}
\includegraphics[scale=0.4]{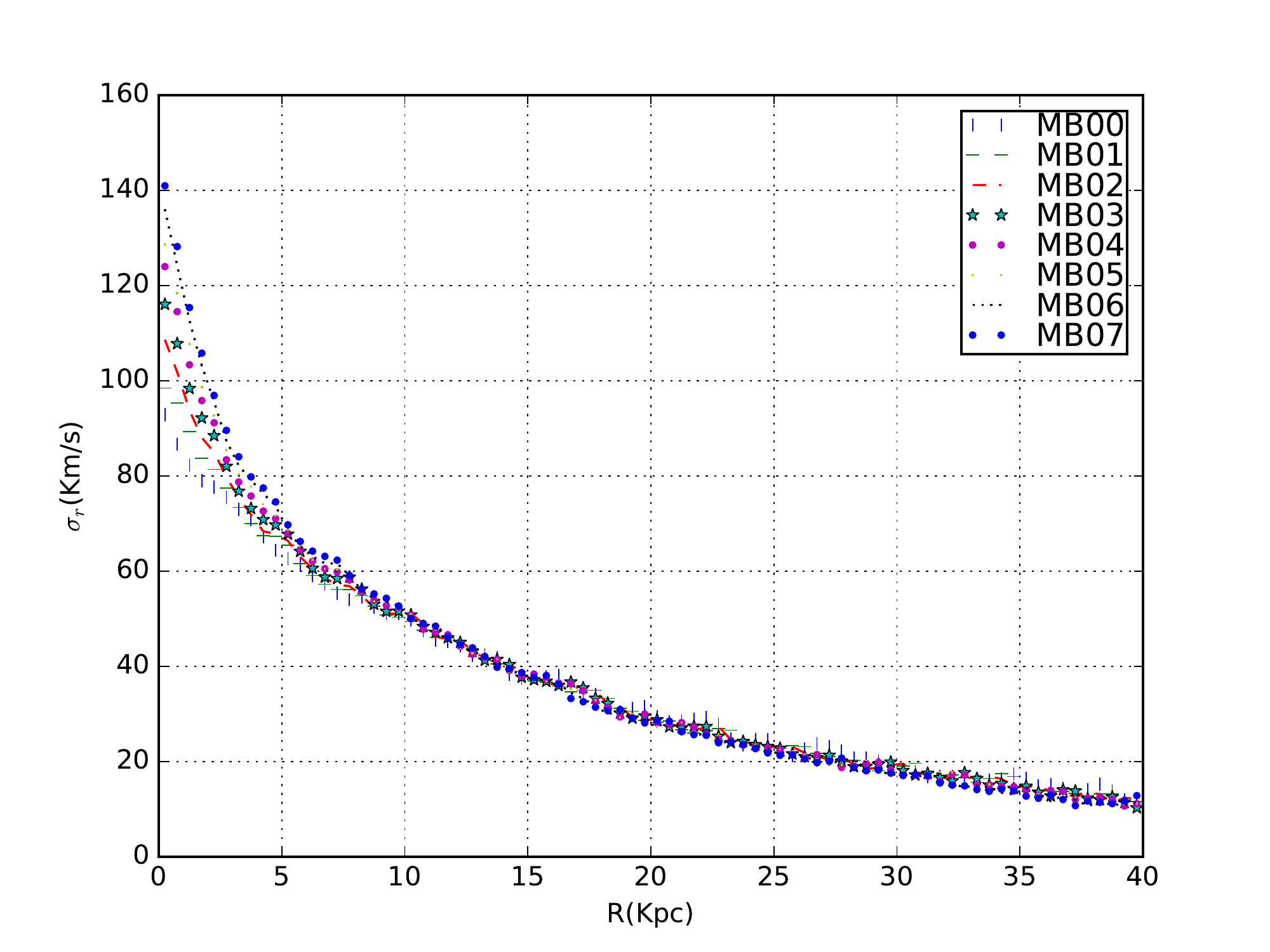}
\caption{Initial radial velocity dispersions for all the MB models}
\label{fig sigma_MB}
\end{figure}

We present here two categories of galaxy models that we denote as MA and MB (here onwards in this article) that have disk, bulge and dark matter halo components. The models MA have relatively smaller disks and higher disk mass surface densities than that of models MB. The purpose of taking these two models is to see the growth and evolution of bar instabilities in both larger and smaller disk galaxies. The galaxy models MA have more concentrated or denser bulges compared to models MB. The details of various parameters in initial models generated is shown in Table \ref{table:Models}. From top to bottom the bulge mass content is increasing in both MA and MB models respectively. Fig. \ref{fig rot_MA} and \ref{fig rot_MB} show initial and final rotation curves, initial surface density and initial toomre parameter with radius for both MA and MB models respectively. Fig. \ref{fig:rotcomp_MA} and \ref{fig:rotcomp_MB} shows the contribution of the different galaxy components to the rotation curves of the galaxy models. Fig \ref{fig sigma_MA} and \ref{fig sigma_MB} show the initial radial velocity dispersion of stellar disk for models MA and MB respectively.

The density profile for a spherically symmetric halo is choosen as
\begin{equation}
\rho_h=\dfrac{M_{dm}}{2 \pi} \dfrac{a}{r(r+a)^3}
\end{equation}   

\begin{table*}
 \caption{Initial Disk Models with increasing bulge masses}
\label{tab:Model Galaxy}
 \begin{tabular}{lcccccccc}
 \hline
 
Models & $\dfrac{M_{B}}{M_{D}}$ & $\dfrac{M_B}{M_T}$ & $\dfrac{M_D}{M_T}$&  $\dfrac{M_H}{M_T}$& $\dfrac{R_b}{R_d}$ & $Q(R_{D})$ &$M_{B}$ & $t_{OP}$\\
   & & & & & & & $[10^{10} M_{\sun}]$ &\\
  \hline
  MA00 &0 & 0&  0.1&0.9 & 0 & 1.077 & 0 &0.180    \\
  MA01 &0.1 &0.01 &0.1&0.89 & 0.174&1.123&0.64 &0.179  \\
  MA02 & 0.2&0.02 & 0.1& 0.88& 0.180&1.171 & 1.28 &0.179  \\
  MA03 & 0.3& 0.03 &0.1 & 0.87&0.186& 1.220 & 1.93 &0.177  \\
  MA04 &0.4 &0.04 & 0.1& 0.86& 0.192&1.269 & 2.55 &0.175  \\
  MA05 & 0.5&0.05 & 0.1& 0.85& 0.198&1.319 & 3.19 &0.176  \\
  MA06 &0.6 & 0.06 & 0.1 & 0.84& 0.204&1.368 & 3.83&0.177  \\
  MA07 &0.7 &0.07 & 0.1 & 0.83&0.210&1.418 & 4.47 &0.177  \\
  \hline
  MB00 &0 & 0& 0.1& 0.9 & 0 & 1.239 & 0 &0.192\\
  MB01 &0.1 &0.01 &0.1 &0.89 & 0.439&1.273 & 1.86&0.193\\
  MB02 & 0.2&0.02 &0.1 &0.88 & 0.447&1.305 & 3.72&0.194\\
  MB03 & 0.3& 0.03 & 0.1&0.87 &0.456& 1.337 & 5.58&0.195 \\
  MB04 &0.4 &0.04 & 0.1&0.86 & 0.465&1.370 & 7.44&0.196\\
  MB05 & 0.5&0.05 &0.1 &0.85 & 0.473&1.401 & 9.30&0.197\\
  MB06 &0.6 & 0.06 &0.1 &0.84 & 0.481&1.433 & 11.16&0.198\\
  MB07 &0.7 &0.07 &0.1 &0.83 &0.490&1.465 & 13.02&0.199\\
  \hline
  \label{table:Models}
   \end{tabular}
\begin{flushleft}
column(1)Model name (2)Ratio of bulge to disk mass (3) Ratio of bulge to total galaxy mass (4) Ratio of disk to total galaxy mass (5) Ratio of halo to total mass (6) Ratio of half mass bulge radius to disk scale length($R_{b}/R_{d}$) (7)Toomre parameter at disk scale length $R_d$(8) Bulge mass (9) Ostriker and Peebles criterian for bar instability $t_{OP}$ explained in section~\ref{criteria} 
\end{flushleft}   

\end{table*}
where a is the scale length of the halo component. This scale length is related to the concentration parameter of the NFW halo having $M_{dm}=M_{200}$ \citep{54a} so that the inner shape of the halo is identical to the NFW halo. Here a and c are related as following
\begin{equation}
a=\dfrac{R_{200}}{c} \sqrt{2[ln(1+c)- c/1+c]}
\end{equation}
where $M_{200}$, $R_{200}$ are virial mass and virial radius for NFW halo respectively. 

The density profile for the disk component has an exponential  distribution in the radial direction and $sech^2$ profile in the vertical direction. 

\begin{equation}
\rho_d=\dfrac{M_d}{4 \pi z_0 h^2} exp\Bigg(-\dfrac{R}{R_d}\Bigg) sech^2\Bigg(\dfrac{z}{z_0}\Bigg)
\end{equation}
where $R_d$ and $z_0$ are the radial scale length and vertical scale length respectively. The values of these parameters are listed in Table \ref{table:Models} for all of our models. 

Finally the bulge component in our models has a Hernquist density profile given by
\begin{equation}
\rho_b=\dfrac{M_{b}}{2 \pi} \dfrac{R_b}{r(r+R_b)^3}
\end{equation}   
where $M_b$, $R_b$ are total bulge mass and bulge scale length respectively. These values are listed in Table \ref{table:Models}.

Table \ref{table:Models} also contains many other parameters like bulge to disk fraction, bulge to total galaxy mass fraction, disk to total galaxy mass fraction, halo to total galaxy mass fraction, Toomre factor at disk scale length ($R_{d}$) and bulge mass. For all the MA and MB type models the total mass of the galaxies is decided by the velocity at virial radius which in our model is equal to 140 Km/s and 200 Km/s respectively. Total galaxy mass for MA galaxy models is equal to 63.8 x$10^{10} M_{\sun}$ and for MB type models it is 186 x$10^{10} M_{\sun}$. Mass content in disk component for all the MA and MB models are 6.38 x$10^{10} M_{\sun}$ and 18.6x$10^{10} M_{\sun}$ respectively. In our models, ratio of particle masses in halo, disk and bulge varies for different models. For MA models, ratio of particle masses goes from 1:1.12:0.22 for MA01 to 1:1.2:1.68 for MA07. For MB models, ratio of particle masses of halo, disk and bulge varies from 1:1.2:0.22 for MB01 to 1:1.20:1.03 for MB07. Both of our models are dark matter dominated and dark matter makes up 90$\%$ of the total galaxy mass. In our models we have chosen the spin parameter for the dark matter halo component  to be 0.035 which is shown to be the most probable value from cosmological simulations \citep{54b}. Bulges in all the models are non-rotating. We have checked that all the models are locally stable as Toomre parameter is greater than 1 for all the models through out the disk. The Toomre factor varies with radius and is given by $Q(r)=\dfrac{\sigma(r)\kappa(r)}{3.36G\Sigma(r)}$. Here $\sigma(r)$ is the radial dispersion of disk stars, $\kappa(r)$ is the epicyclic frequency of stars and $\Sigma(r)$ is the mass surface density of the disk.    

\subsection{Simulation Method}
After all the initial galaxy models are prepared we evolve these models in isolation with Gadget-2 code \citep{56}. We evolved these galaxies up to 9.78 Gyr to check secular evolution of the disks in our models. This code uses various types of leapfrog methods for time integration. It uses the Tree method \citep{56b} to compute gravitational forces between different particles. The opening angle for the tree is chosen as $\theta_{tot}=$0.4. The softening length for halo, disk and bulge components has been chosen 30, 25 and 10 pc respectively. We mention our results in terms of code units. Both GalIC and Gadget-2 code have unit mass equal to $10^{10}$  $M_{\sun}$, unit distance is 1 kpc, unit velocity is 1 km/s.

The introduction of concentrated bulges increases the frequency of 2 body interactions in our simulations and affects the conservation of angular momentum of the galaxy. This means that the force calculation accuracy plays a important role in determining position and velocities of the particles and therefore angular momentum conservation. We have done many test simulations to determine the most important parameters for conserving the angular momentum. We found that reducing the softening length does not help much in conserving the angular momentum.
Instead for our current simulation, reducing the time step of integration ($\eta$) and the force accuracy parameter over the evolution time period are the most important parameters for angular momentum conservation\citep{klypin2009}. We haves used the values $\eta< = $~0.15 and force accuracy parameter $< =$~0.0005 in most of the simulations. As a result in all of our models, the angular momentum is conserved to within 1 $\%$ of the initial value.

\subsection{Bar strength and Pattern Speed}
Bar strength has been defined in different ways in the literature   \citep{13a,15a1}. In our study for defining  bar strength we have used the mass contribution of disk stars to the m=2 fourier mode.

\begin{equation}
a_2(R)=\sum_{i=1}^{N}  m_i \cos(2\theta_i)\\
b_2(R)=\sum_{i=1}^{N} m_i \sin(2 \theta_i)
\end{equation}

where $a_2$ and $b_2$ are defined in the annulus around the radius R in the disk, $m_i$ is mass of $i^{th}$ star, $\theta_i$ is azimuthal angle. We have defined the bar strength as 
\begin{equation}
\dfrac{A_2}{A_0}=max \Bigg(\dfrac{\sqrt{a_2 ^2 +b_2 ^2}}{\sum_{i=1}^{N} m_i}\Bigg) 
\end{equation}

We calculate the pattern speed($\Omega_B$) of the bar by measuring change in phase angle  $\phi=\dfrac{1}{2}\tan^{-1}\bigg(\dfrac{b_2}{a_2}\bigg)$ of the bar which is calculated using the fourier component in the annulus corresponding to maximum bar strength. We use annular regions of size 1 kpc for disk particles only. 

\subsection{Angular Momentum Calculation}
We measured the  angular momentum of the different components of a galaxy separately; disk, bulge and halo. Angular momentum of a particle is calculated using the product of its particle mass, radial distance from galactic center and circular velocity. In this paper we plot time evolution of total angular momentum of each component in the galaxy.

\section{Results} \label{Results}

\subsection{Evolution of Bar Strength}
\begin{figure}
\includegraphics[scale=0.4]{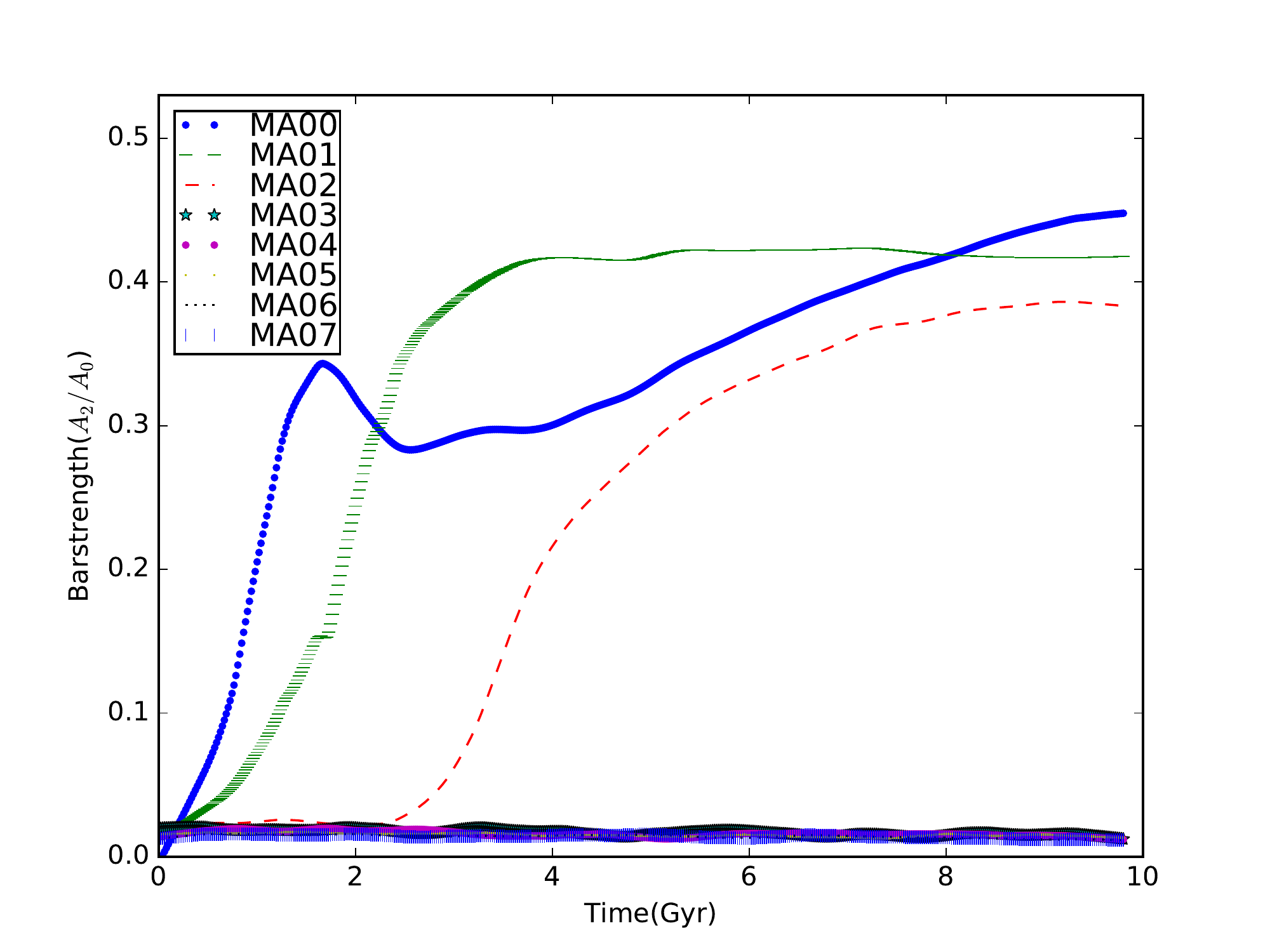}
\caption{Evolution of bar strength with time for all MA models.}
\label{figure:BS_MA}
\end{figure}

\begin{figure}
\includegraphics[scale=0.4]{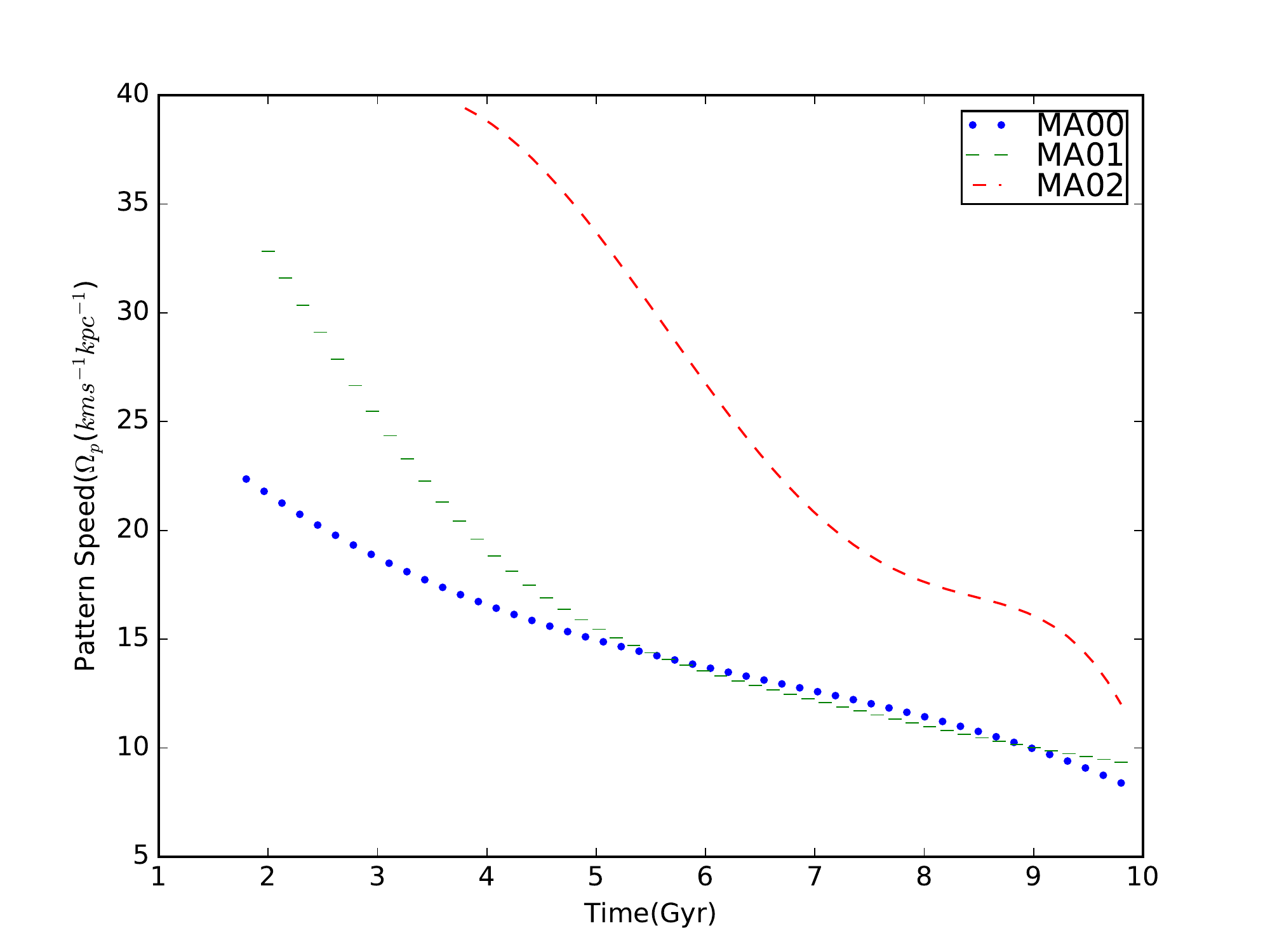}
\caption{Pattern speed evolution with time  of all bar forming MA models }\label{PS_MA}
\end{figure}

\begin{figure}
\includegraphics[scale=0.32]{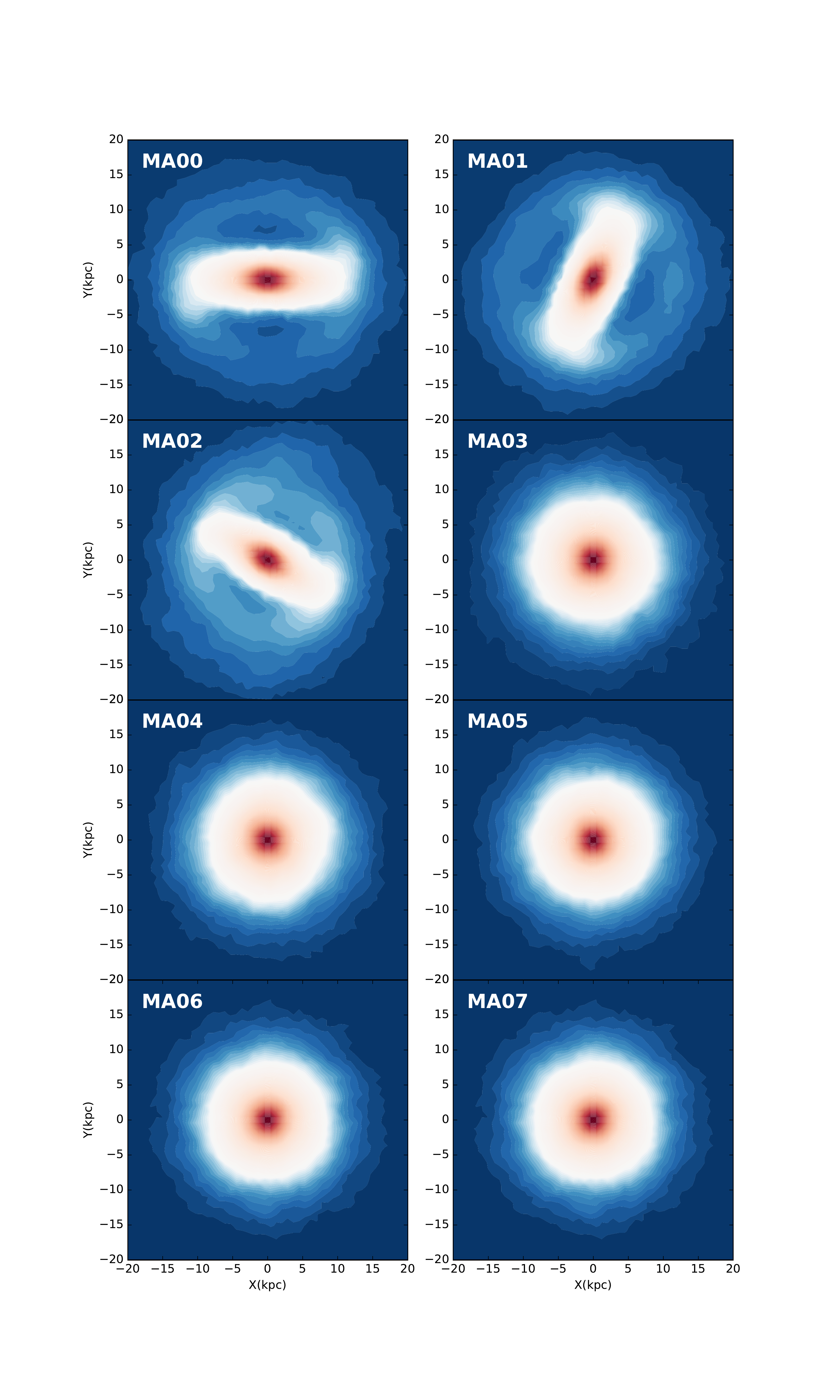}
\caption{X-Y cross section of all the MA models at 9.78 Gyr  }\label{fig MA}
\end{figure}

We tracked bar formation and evolution up to 9.78 Gyr in all of the models from MA00 to MA07. Figure \ref{figure:BS_MA} shows the evolution of bar strength for these models. The X-Y face-on view of all the MA models after 9.78 Gyr is shown in Figure~\ref{fig MA}. We see that the model without bulge (MA00) grows to peak bar strength  around 1.5 Gyr during dynamical evolution and it's bar strength decreases rapidly before secular evolution phase. In secular evolution it's bar strength increases gradually with time. In the model with low mass bulges (MA01) the bar grows to peak strength around \textbf{3 Gyr} and sustains its strength during secular evolution phase. Further with increasing bulge masses in model MA02, bar instability sets in later and reaches peak value around \textbf{7 Gyr.} In the later models MA03, MA04, MA05, MA06 and MA07 that have bulge to disk  fraction > 0.3 bar type of instability is suppressed completely due to the presence of massive bulge. 

\begin{figure}
\includegraphics[scale=0.32]{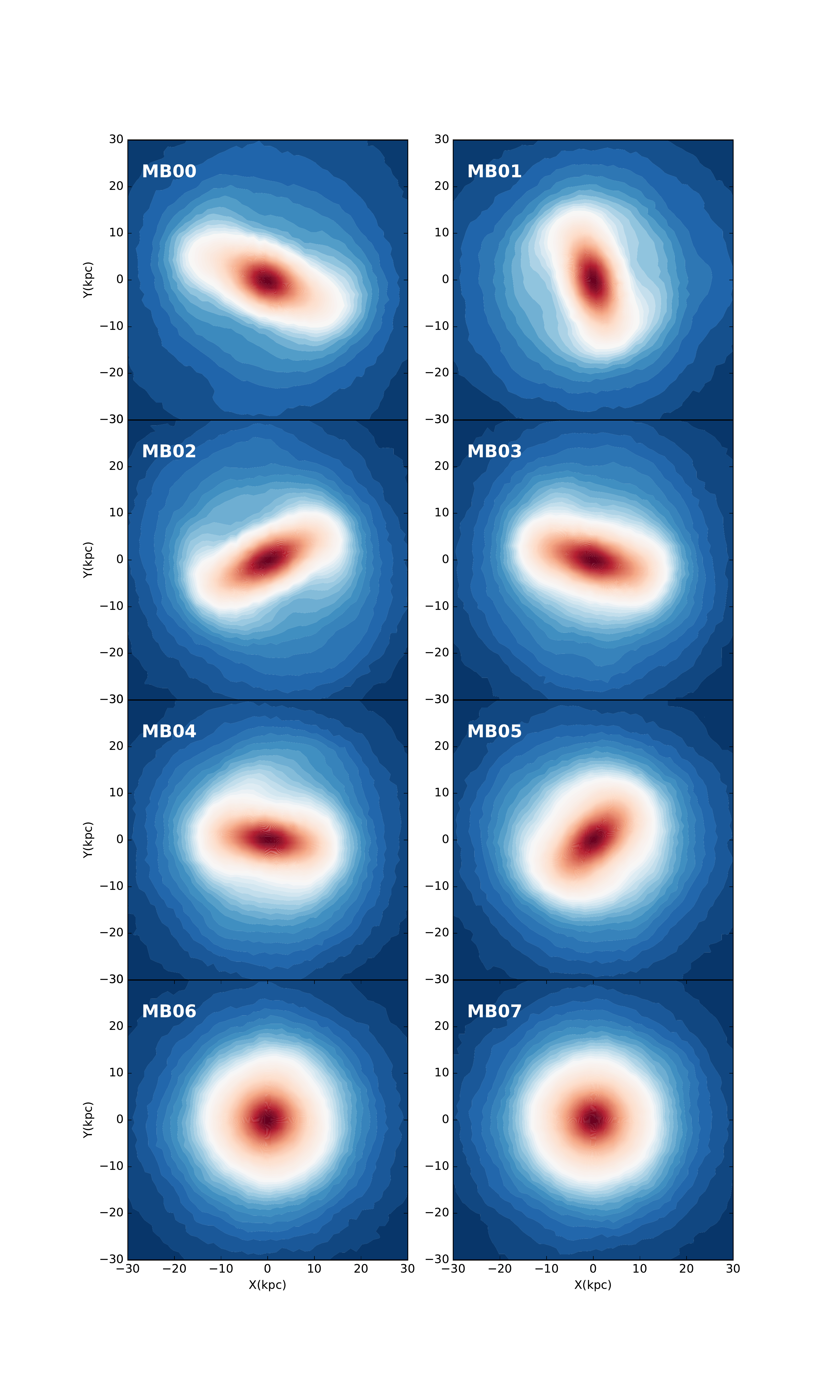}
\caption{X-Y cross section of all the MB at 9.78 Gyr  }\label{fig MB}
\end{figure}

\begin{figure}
\includegraphics[scale=0.4]{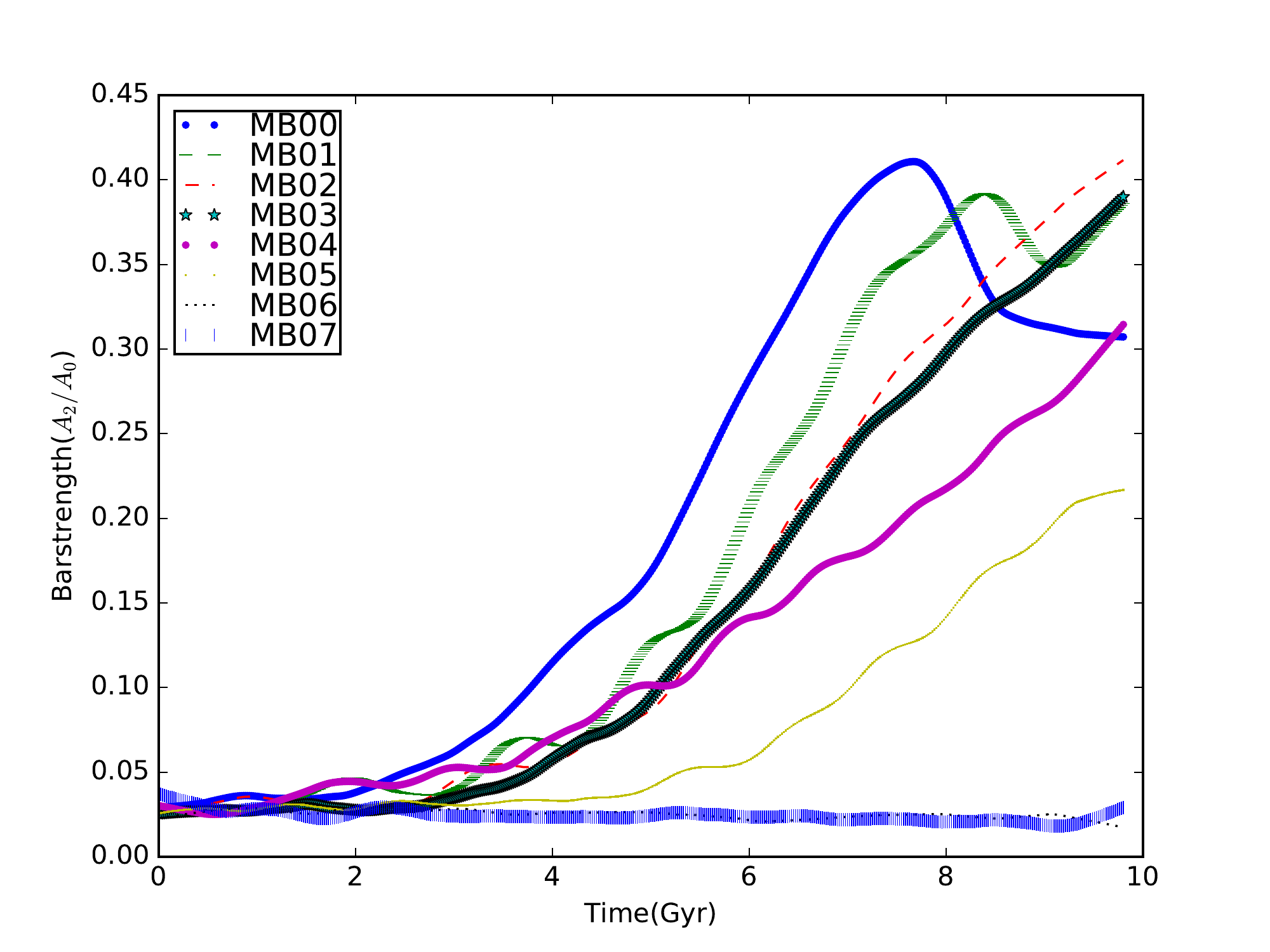}
\caption{Evolution of bar strengths with time for all MB models.}
\label{figure:BS_MB}
\end{figure}

Figure~\ref{figure:BS_MB} shows evolution of bar strength for the models MB00 to MB07, that have less concentrated bulges and lower disk surface densities compared to models MA00 to MA07. As in the previous models, the simulation is run for 9.78 Gyr. The main difference in the bar evolution of models MA and MB is that the bar forms much later in model MB. This is due to the lower mass surface density $\Sigma$ of the disk which leads to a lower disk self gravity and hence instabilities take longer to develop. We find that the model with no bulge (MB00) shows bar type instability, for which bar strength peaks at around \textbf{7.5 Gyr} and the bar gets weaker with further evolution. On introduction of a bulge in our MB models we see that the bar strength shows nonlinear trends as a function of bulge to disk fraction. For small bulge to disk ratios of $~ $0.1- 0.3, the bar triggering time scale increases with bulge mass and the peak bar strength remains almost the same. As we increase bulge fraction from 0.4 to 0.6 we see that the bar formation time scale increases and peak strength reduces for the time we evolve our model. For models with bulge to disk fraction 0.7, a bar does not  form at all. The X-Y face on views of all the MB models after 9.78 Gyr is shown in Figure \ref{fig MB} 

\subsection{Evolution of Bar Pattern Speed}
We show the evolution of bar pattern speed ($\Omega_p$) with time in Figure~\ref{PS_MA} for all MA models which form bars. In this plot we do not have models MA03 to MA07 because their disks do not form bars. It is very clear that the disks that have more massive bulges have bars with higher $\Omega_p$. Also, in all the models, the $\Omega_p$ decreases with time and the rate of decrease in pattern speed increases with increase in bulge mass fraction.

\begin{figure}
\includegraphics[scale=0.4]{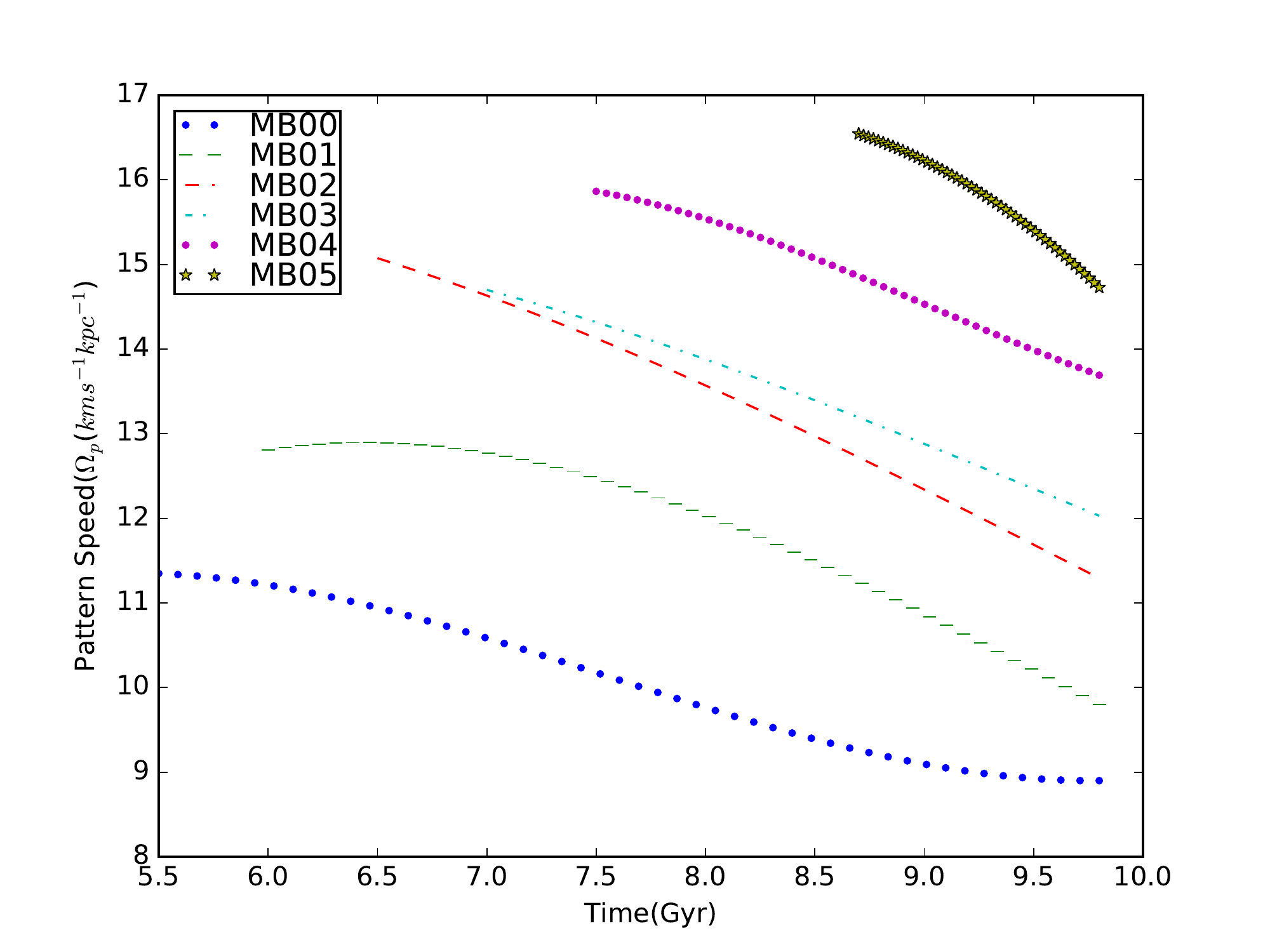}
\caption{Pattern speed evolution with time  of all bar forming MB models }\label{PS_MB}
\end{figure}

The variation of bar pattern speed with time for all MB models which form bars, is shown in Figure~\ref{PS_MB}. Note that we can plot $\Omega_p$ only after the bars form and start growing. So this plot excludes models MB06 and MB07 which do not form bars. As in models MA, as the bulge fraction increases (from MB00 to MB05), the $\Omega_p$ value is higher. Thus, bar rotation is faster for bulge dominated galaxies, whatever be the bulge concentration or disk surface density. This is because the increase in bulge mass makes the inner disk potential deeper resulting in larger angular velocities. In all the bar forming models, $\Omega_p$ shows little variation for several Gyr until 8 to 9~Gyr, after which $\Omega_p$ decreases sharply with time. We interpret that this may be related to the thickening  of the bar as the bar growth peaks after this time period.

\subsection{Angular Momentum Exchange}

\begin{figure*}
\includegraphics[scale=0.37]{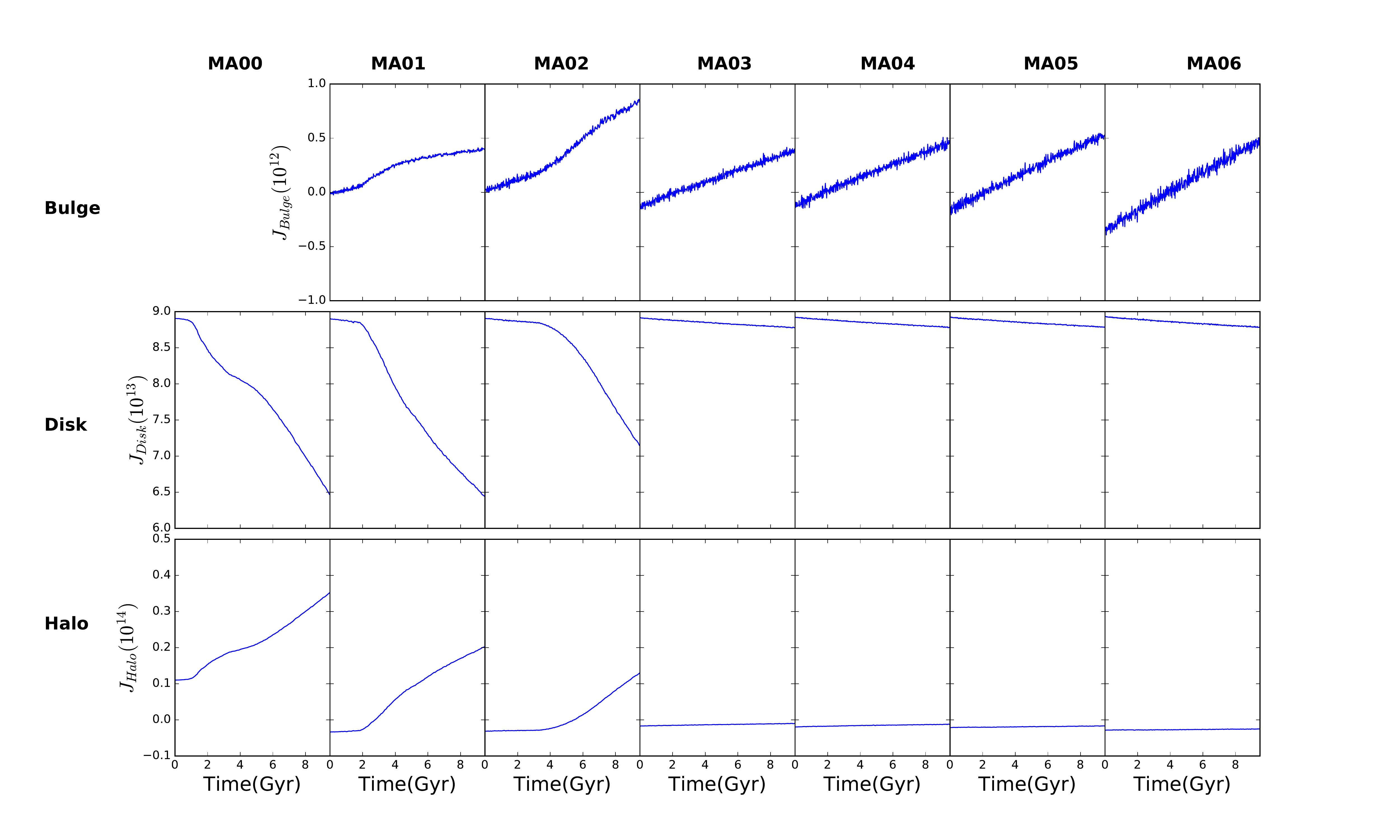}
\caption{Total Angular Momentum for all the components(Bulge, Disk and Halo) of all MA models}
\label{fig Total_AngularMomentum_MA.eps}
\end{figure*}

\begin{figure*}
\includegraphics[scale=0.37]{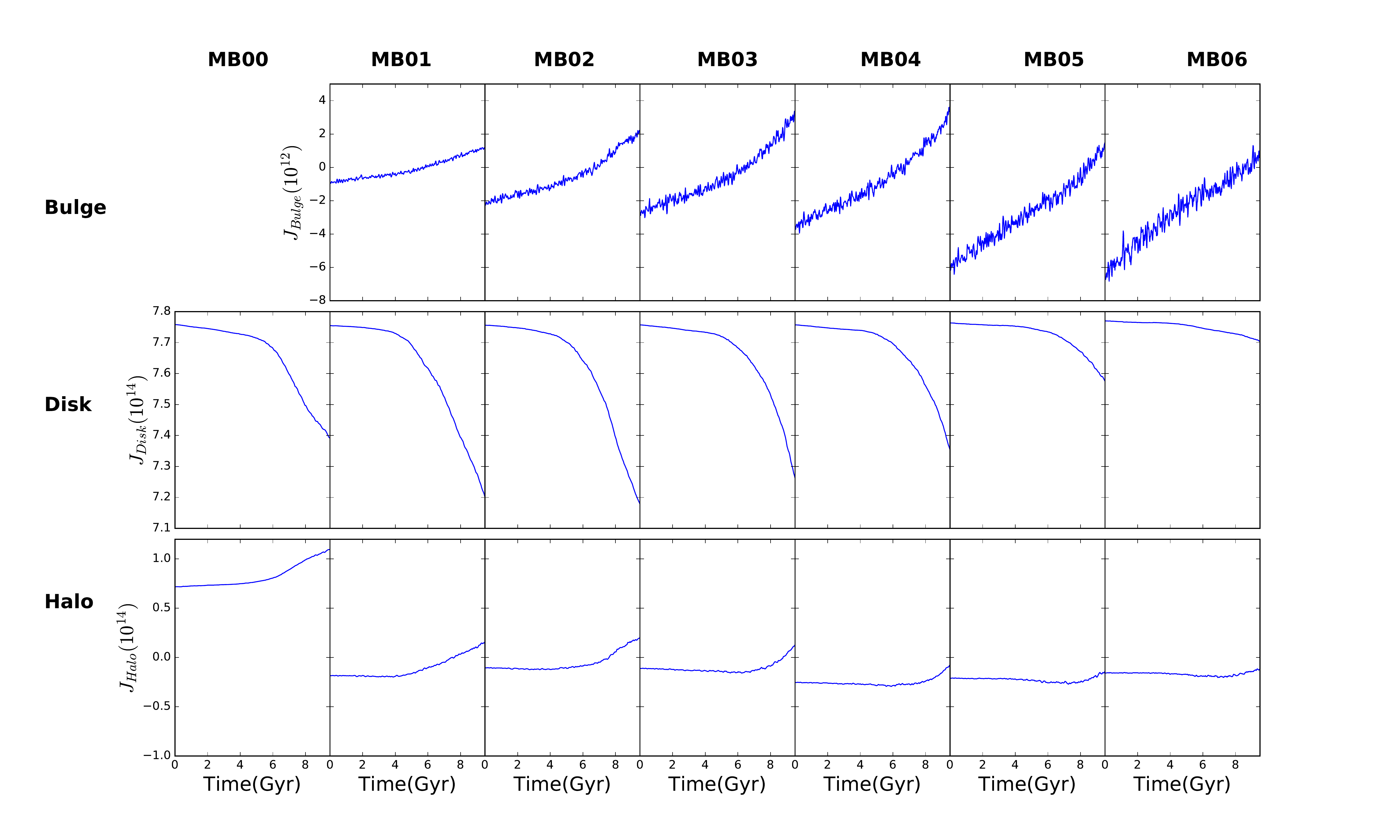}
\caption{Total Angular Momentum for all the components(Bulge, Disk and Halo) of all MB models}
\label{fig Total_AngularMomentum_MB.eps}
\end{figure*}

Several studies have shown that angular momentum exchange between the disk and halo plays a vital role in the formation and evolution of bars \citep{15a1,15a,Martinez.2006,17,18}.
We see this in both MA and MB type of models in our simulations. In this sub-section we discuss the importance of angular momentum transport between the disk, halo as well as bulge in our models. We have also plotted the total angular momentum exchange with time for the individual galaxy components in Figure~\ref{fig Total_AngularMomentum_MA.eps} and Figure~\ref{fig Total_AngularMomentum_MB.eps} for MA and MB type models respectively.  
 
\subsubsection{MA Models : High density bulges and disks} 

The overall change in angular momentum for bulge, disk and halo components of MA models are shown in Figure~\ref{fig Total_AngularMomentum_MA.eps}. 
We see that disk component loses total angular momentum by huge amount in bar forming models compare to models which does not show bar instability.
We also see that angular momentum loss rate for the disks increases sharply when the bars reach their peak strength, the time scale for which increases with increase in bulge mass.We can see that the total angular momentum gain is maximum for the halo component and is around 15 to 20 times that of the bulge component. \\ 
\indent  We also note that the total angular momentum of the bulge component first increases with increase in bulge mass for models MA00, MA01 and MA02 which show bar instabilities. Further increase in bulge mass in model MA03 leads to decrease in bulge angular momentum. After this, an  increase in bulge mass (Model MA04 onwards) results in a slow increase in the bulge total angular for all the bulges and could represent the slow spin up of bulges. \\                   
\indent For the halo component, the total angular momentum exchange is
maximum for the MA00 model which does not have a bulge. On introduction
of bulge for models MA01 and MA02, the halo component gains angular momentum after the bar gains its peak strength around 2 Gyr and 4 Gyr respectively. For models MA03 onwards where the  bar instability is not triggered and the bulges mass increases, the angular momentum exchange becomes very small.

\subsubsection{MB Models : low density bulges and disks}

  In all the MB models, the bar forms after 6~Gyr which is much later than in the MA models, mainly because the the disk mass surface density is much lower. Figure~\ref{fig Total_AngularMomentum_MB.eps} shows total angular momentum change for bulge, disk and halo components of all the MB models. We see that disk component loses angular momentum by large amount in bar forming models. Similar to MA models here also the rate of decrease in angular momentum correlates with the bar strength peaking time scale which is shown in  Figure~\ref{figure:BS_MB}. Total loss of angular momentum by disk component decreases with increase in bulge mass as we go from MB01 to MB07. In these models we see that both halo and bulge components gain angular momentum. Also, the total angular momentum gain is maximum for the halo component and is around 5 to 40 times that of the bulge component.
   \\
\indent We see that increase in total angular momentum of bulge increases with increase in bulge masses. It can also be seen clearly that rate of increase of angular momentum for bulge components increase after bar strengths peak which varies in all the models. \\
\indent The rate of gain in angular momentum by the halo component also start increasing only after the bar gets its peak strength. Further total change in angular momentum of halo component decreases with increase in bulge mass. This clearly shows that a massive bulge component delays bar formation by curbing angular momentum transport from disk component to halo component.

\subsection{Bar Instability criteria} \label{criteria}
Earlier works have shown that cold stellar disks \citep{12a,12b,15a1} with low velocity dispersion can become bar unstable within a few gigayears of evolution. Ostriker and Peebles have given a criteria that if the ratio of rotational kinetic energy to potential energy of a disk exceeds 0.14 (t$_{OP}~>~0.14$), then the galaxy disk is bar unstable \citep{24}. In all of our models(MA and MB), we see that all of the t$_{OP}~>~0.14$. This suggests that all of our models should have formed bars as the disk is cold enough to become bar unstable. We show the values of t$_{OP}$ for all the models in Table~1. We find that models MA become bar unstable when t$_{OP}~>~0.177$ and models MB become unstable when t$_{OP}~>~0.197$. Thus the t$_{OP}$ does not appear to be constant in the different disk models and depends on the disk surface mass density as well as the bulge mass concentration. We suggest that this shift in Ostriker and Peebles criteria to higher values is due to live/spinning nature of halo \citep{17} and bulge components.

Toomre (1981) has shown that bar instabilities can occur
through the feedback mechanism during swing amplification. This
feedback mechanism is cut down in the presence of ILR resonances, that do not allow the waves to propagate and hence bars should not form in the presence of ILR's. However, we do not find that the presence of ILRs in our simulation is a deciding criteria for bar formation.

\begin{figure}
\includegraphics[scale=0.45]{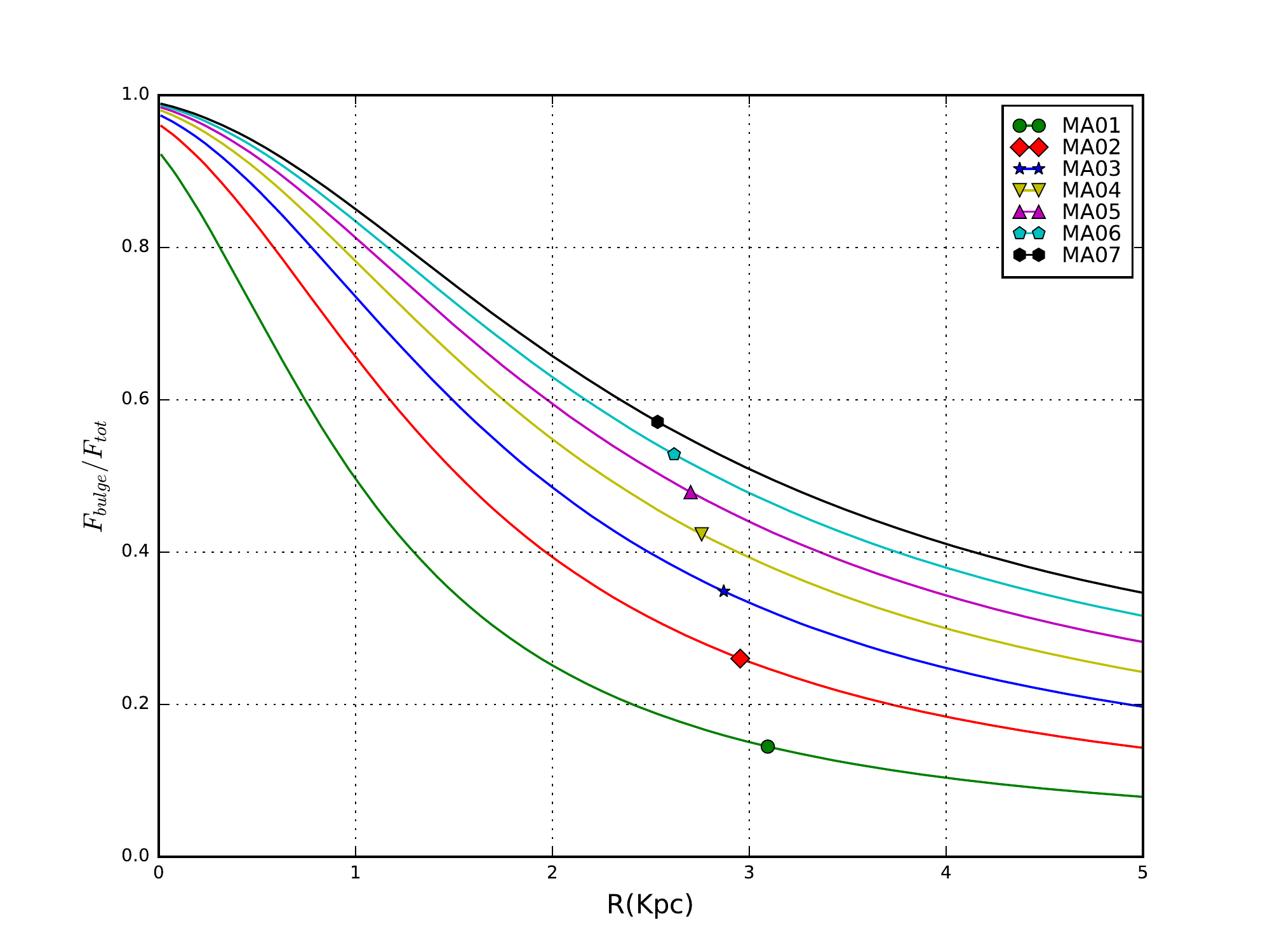}
\caption{Fractional bulge force at disk scale length for MA Models( various shape dots)}
\label{fb_MA.eps}
\end{figure}

\begin{figure}
\includegraphics[scale=0.45]{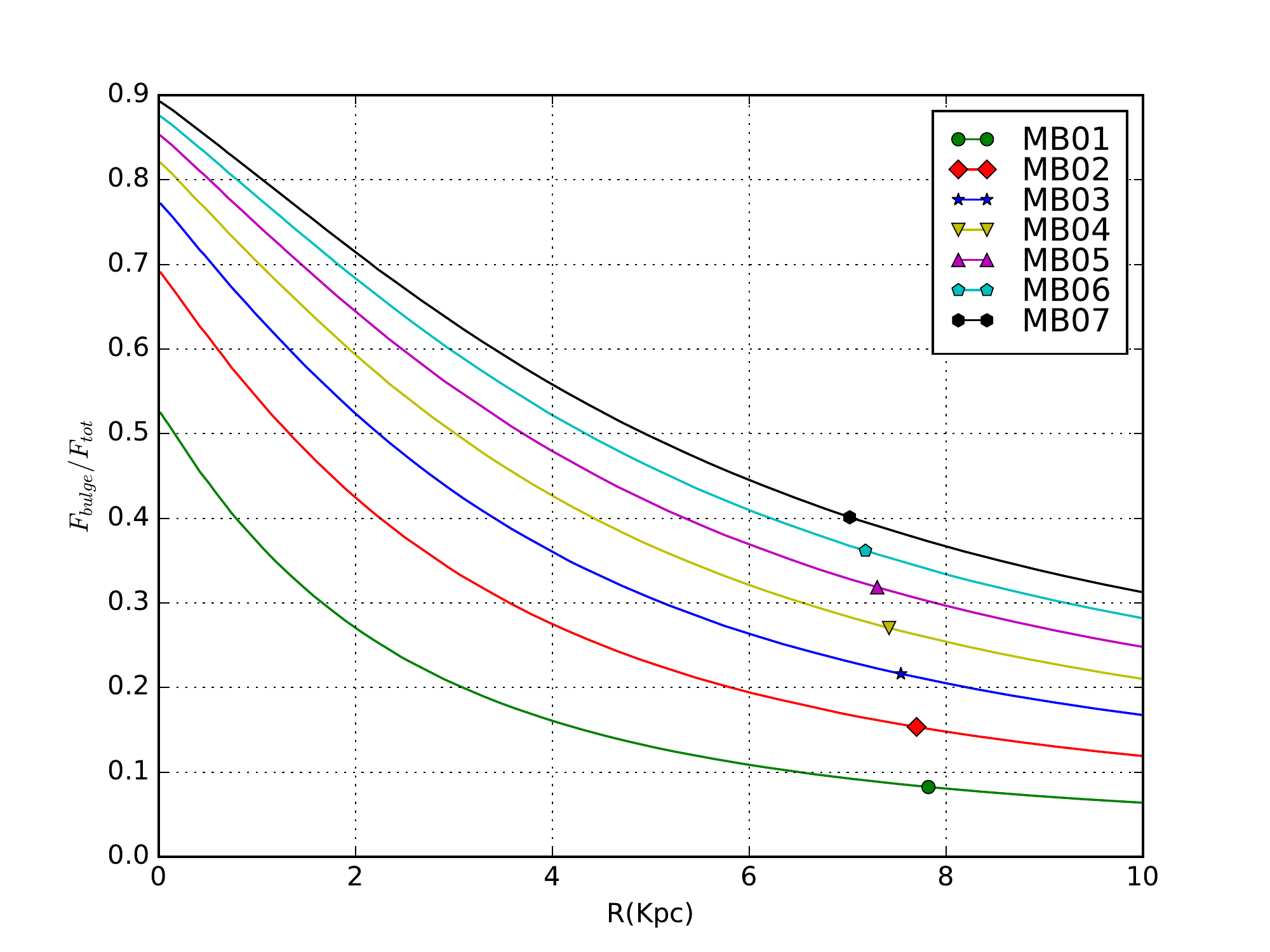}
\caption{ Fractional bulge force at disk scale length for MB Models( various shape dots)}
\label{fig fb_MB.eps}
\end{figure}

\begin{figure}
\includegraphics[scale=0.45]{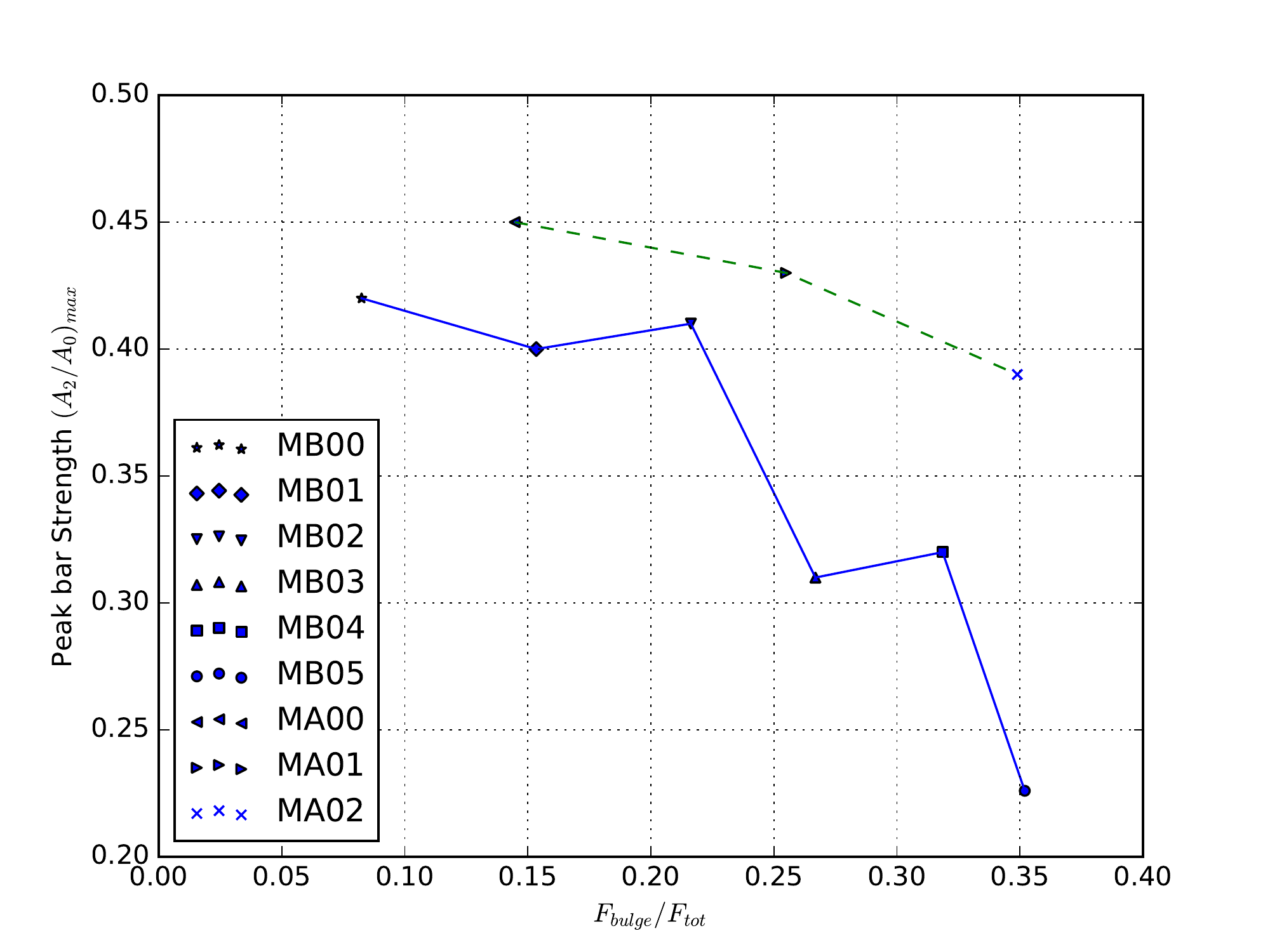}
\caption{ Peak bar strength with respect to fractional bulge force for all the bar forming MA(dashed line) and MB models(solid line).}
\label{fb_different.eps}
\end{figure}

As we know bulges are the prominent mass component in the centres
of galaxies. Hence, they contribute significantly to disk rotation in the central regions. We have found that all the bar forming models have fractional bulge to total radial force values, $F_{b}/F_{tot}$ at the disk scale radius $R_d$, to have values of less than 0.35. Here $F_{b}=V^2_{b}/R_d$ and $F_{tot}=V^2_{tot}/R_d$; $V_{b}$is velocity due to bulge component and $V_{tot}$ is velocity due to bulge,disk and halo components. We have plotted this ratio $F_{b}/F_{tot}$ with radius in Figures~\ref{fb_MA.eps} and \ref{fig fb_MB.eps} for the models MA and MB. The cut off for fractional bulge force is shown clearly in Figure ~\ref{fb_different.eps}. The dots of different shapes represent the values of $F_{b}/F_{tot}$ at disk scale lengths and the value is never larger than 0.35. Hence, our work shows that bars can form in disk galaxies only when $F_{b}/F_{tot}~<~0.35$. This happens because velocity dispersion of disk stars increases as result of increase in radial force due to increasing mass of bulge component. We thus propose that this is a new criteria for bar formation in terms of radial force due to the bulge mass in the disk. This criteria gives clear idea about the amount of bulge force that prevents bar formation, regardless of the bulge mass or concentration.

\subsection{Boxy Bulges in the models}
\begin{figure*}
\includegraphics[scale=0.65]{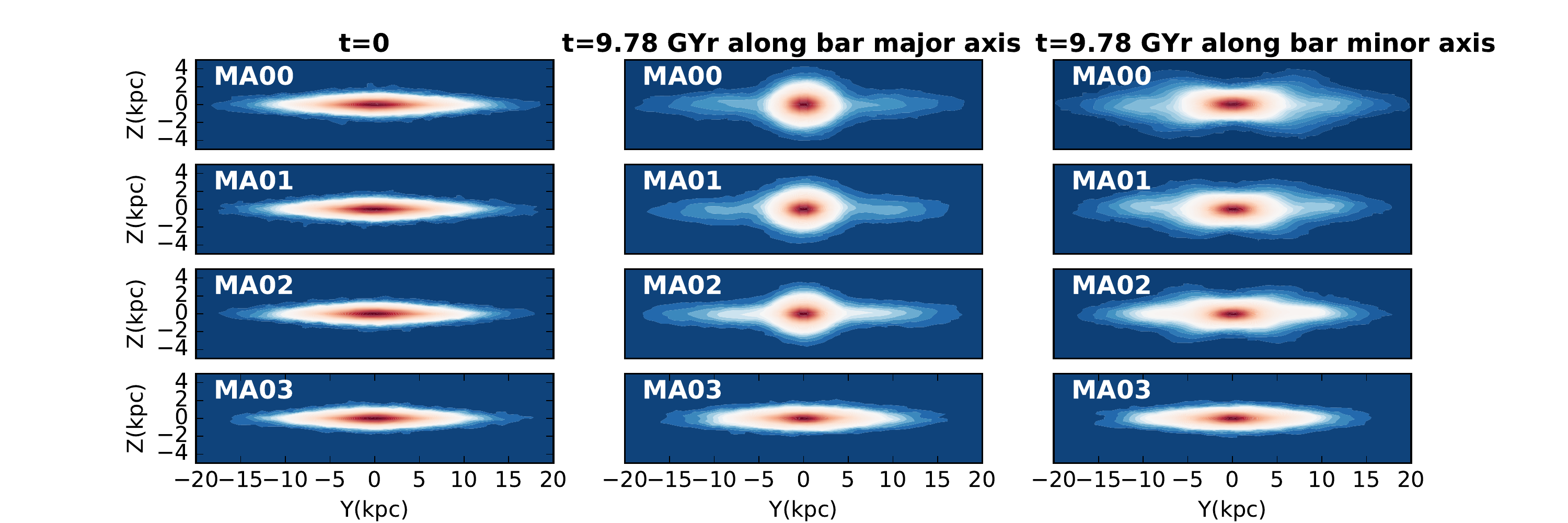}
\caption{First column shows edge on(YZ cross section) of MA models at t=0; second and third column shows 2 different edge on view at t=9.78 Gyr} 
\label{Final_imagesyz_MA}
\end{figure*}

\begin{figure*}
\includegraphics[scale=0.435]{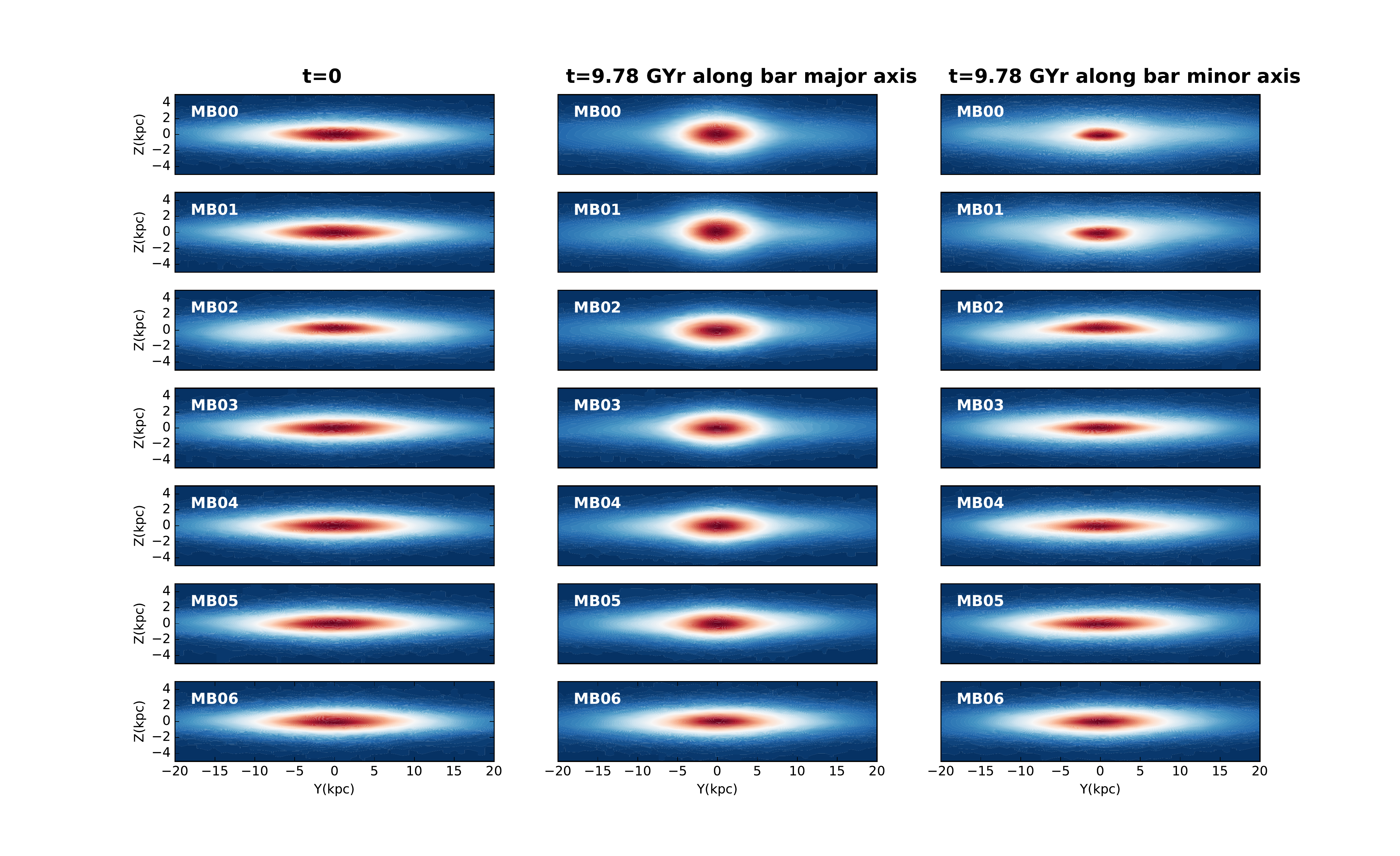}
\caption{First column shows edge on(YZ cross section) of MB models at t=0; second and third column shows 2 different edge on view at t=9.78 Gyr} 
\label{Final_imagesyz_MB}
\end{figure*}

Figure~\ref{Final_imagesyz_MA} and Figure~\ref{Final_imagesyz_MB} show the peanut/box(P/B) shape features at the end of evolution in both the MA and MB models respectively in the y-z planes at both start and end of the simulations. We see that only the bar forming models show P/B feature. Their origin is due to heating of bar through vertical resonances \citep{33}. P/B features are stronger for MA models than MB models. This is because the bar forms earlier (t$\sim$2Gyr) in MA models and this provides sufficient time for bar heating during the secular evolution phase. While in MB models the bar forms much later (t$\sim$6 Gyr) and it does not  have enough time to secularly evolve. We find that there is a weak correlation of these P/B feature and the classical bulge mass fraction in both the MA and MB models. 

\section{Implications Of Our Results}\label{Discussion}
Our simulations show that the formation of bar instabilities changes 
with bulge mass and bulge concentration. Both effects can be explained by the criterion $F_{b}/F_{tot}~<~0.35$ for bar formation, as described in the previous section. The cut off implies that angular momentum transfer among galaxy components gets damped due to increase in velocity dispersion of stars due to introduction of massive bulges. This means that galaxies with strong bulges will not form bars easily in their disks, perhaps only during interactions with other galaxies as well. Or in other words, strong bulges make disks extremely stable. Such strong bulges may form during galaxy formation epochs, in which case the galaxies do not form bars easily or during secular evolution when gas is driven into the nuclear region by bars or spiral arms leading to the build up of central mass concentrations \citep{58,59,60}.  

 Our bulge related bar instability criterion can be easily related to observations in the following way. For a given galaxy image in the near-infrared (which is a waveband that traces the main stellar content of galaxy), a bulge-disk-bar image decomposition will yield the luminosity of the individual components (L$_ {bulge}$, L$_{disk}$, L$_{bar}$). Further force due to bulge component can be calculated through mass of bulge ($M_{bulge}$), which can be obtained by multiplying M/L ratio with L$_ {bulge}$. Rotation curve($V_{tot}$) and disk scale length of the galaxy can be used to obtain total force due to galaxy at disk scale length. Hence, our criterion can be written in observable quantities as:   
 \begin{equation}
     \dfrac{F_b}{F_{tot}} = \dfrac{GM_{bulge}}{R_d V^2_{tot}}
\end{equation}
 
This effect of strong bulges can be clearly seen in observations of bulge dominated disk galaxies. Bars are not common in S0 galaxies that have classical bulges and bulge-disk decomposition for S0's reveal a bulge to total luminosity (B/T) value equal to 0.35 \citep{57}. This value is similar in nature to our criterion of $F_{b}/F_{tot}~<~0.35$. Studies of late type spiral galaxies indicate that bulge dominated spirals have a smaller fraction of barred galaxies compared to disk 
dominated ones \citep{2}. However, later studies with Galaxy Zoo have shown that bars tend to form in disks with stronger bulges \citep{4}. A more rigorous study of B/T values for a large sample of barred galaxies is required to fully test our criterion.
Another important implication of our study is that bars do form in dark matter dominated disks. Our models MB are similar to large LSB galaxies that contain massive dark matter halos which is supposed to make disks stable against bar formation as predicted by classical theory   \citep{12a,24}. We see bar instability in our models despite the massive dark matter halos; this is similar to the observations of low surface brightness galaxies (e.g. UM~163) \citep{26}. This can be explained by the live and rotating nature of halos as seen in our study and others in the literature \citep{17,18}. In our case we have assumed a constant halo spin value equal to 0.035.     
 
Another important implication of this study is the continuous gain in angular momentum by the bulge. We see that for the models that do not form bars, there is no angular momentum exchange with the halo, as shown in Figures \ref{fig Total_AngularMomentum_MA.eps} and \ref{fig Total_AngularMomentum_MB.eps}. But the bulge component always
gains angular momentum irrespective of bar formation in all the
models. The amount of angular momentum absorbed by the bulges is
more in the case when the disk forms bars.

\section{Summary}\label{summary}
In this paper we have used N-body simulations to understand the effect of bulge mass and bulge concentration on bar formation and evolution in disk galaxies. We have evolved two models; models MA have dense bulges and disks of relatively high surface mass density $\Sigma$ and models MB have less dense bulges and low $\Sigma$ values compared to MA. We vary the bulge mass in similar steps for both models.\\
{\bf 1.} For models with concentrated bulges (MA) we notice that a bar forms earlier for bulgeless galaxies compared to those with bulges. The delay in bar formation increases as the bulge to disk mass fraction  increases. No bar forms when the bulge to disk mass fraction is $>$ 0.3.\\
 {\bf 2.} For less dense bulge models (MB), the bar formation time scale increases with bulge mass up to bulge to disk mass fraction equal to 0.5. No bar forms when the bulge to disk mass fractions is $>$~0.6.
 \\ {\bf 3.} Our simulations show that bars are faster for models with massive bulges. The rate of decrease in pattern speed of the bar increases with introduction of massive bulges in our galaxy models.\\  
{\bf 4.} We see that in both MA and MB type models, the introduction of more massive bulges makes the disk stellar dispersion higher in the inner disk region. Hence massive bulges do not allow overall disk-halo angular momentum exchange which in turn makes the disk bar stable. The fraction of bulge to disk mass ratio required to make a disk bar stable varies with the density of bulge. In our study we see that it has a value of 0.3 for the dense bulge models (MA) and 0.6 for the low density bulge model (MB).\\
 {\bf 5.} We put a limit on bar formation in disks with bulges by introducing a quantity B which is the ratio of radial force due to bulge and disk components (B~=~$F_{b}/F_{tot}$) at disk scale length ($r_{d}$). We find that bars do not form for B$>$0.35. This is because the velocity dispersion of disk stars becomes large enough to inhibit bar type instability. This criterion does not depend on bulge concentration but takes into account the radial force due to a bulge that prohibits bar formation. It is similar to earlier values derived for disks with spherical halos  \citep{46}.\\ 
{\bf 6.} Rate of angular momentum transfer to bulge and halo components from disk is more for dense bulges (MA) than less dense ones (MB). \\
{\bf 7.} Bulge component always gains total angular momentum of the same order irrespective of bar formation in any of the simulated models, while the gain in total angular momentum for halo component dampens by huge amount for bar stable models compare to bar unstable ones.

\section*{Acknowledgements}
\addcontentsline{toc}{section}{Acknowledgements}

We thank Denis Yurin and Volker Springel for providing GalIC code which we used for generating initial conditions of galaxy. We thank the authors of the Gadget-2 code \citep{56,61} which we have  used for this study. We are grateful to HPC facility "HYDRA" at Indian Institute of Astrophysics, Bangalore where we ran our simulations. We are also thankful to Lukas Konstandin and Francoise Combes for helpful discussion for this work. We also thank Juntai Shen for suggesting better way to present bar formation critera. At last we thank the anonymous referee for the useful comments which helped in improving the content of this article.






\bsp	
\label{lastpage}
\end{document}